\documentclass[aps,pre,10pt,showpacs,floatfix,onecolumn,nofootinbib,a4paper]{revtex4-2}
\usepackage{amssymb,amsmath,amsthm,mathtools}
\usepackage{bm}
\usepackage{mathrsfs}
\usepackage{hyperref,bookmark}
\usepackage{graphicx}
\usepackage{epsfig,color}
\usepackage{mathrsfs}
\usepackage{verbatim}
\usepackage{url}
\usepackage{subcaption}
\usepackage{float}
\usepackage[greek,english]{babel}
\usepackage{dcolumn}

\newcommand{\tr}{\operatorname{tr}}
\newcommand{\dd}{\operatorname{d}\!}

\newcommand{\diver}{\operatorname{div}}
\newcommand{\curl}{\operatorname{curl}}
\newcommand{\arccot}{\operatorname{arccot}}

\newcommand{\n}{\bm{n}}
\newcommand{\e}{\bm{e}}
\newcommand{\nper}{\bm{n}_\perp}
\newcommand{\normal}{\bm{\nu}}
\newcommand{\body}{\mathscr{B}}
\newcommand{\region}{\mathscr{R}}
\newcommand{\boundaryR}{\partial\region}
\newcommand{\free}{\mathscr{F}}

\newcommand{\Req}{R_\mathrm{e}}

\newcommand{\p}{\bm{p}}

\newcommand{\framec}{(\e_x,\e_y,\e_z)}

\newcommand{\conf}{\mathsf{S}}

\newcommand{\Fa}{F_\alpha}
\newcommand{\vae}{\varepsilon}

\begin{document}
\latintext

\title{Shape Bistability in Squeezed Chromonic Droplets}
\author{Silvia Paparini}
\email{paparinisilvia@gmail.com}
\author{Epifanio G. Virga}
\email{eg.virga@unipv.it}
\affiliation{Dipartimento di Matematica, Universit\`a di Pavia, Via Ferrata 5, 27100 Pavia, Italy}

\date{\today}

\begin{abstract}
	We study droplets of chromonic liquid crystals squeezed between parallel plates inducing degenerate tangential anchoring on the nematic director. In the coexistence regime, where droplets in the nematic phase are at equilibrium with the surrounding melt, our two-dimensional theoretical model predicts a regime of shape bistability for sufficiently large bipolar droplets, where \emph{tactoids} (pointed shapes) and \emph{discoids} (smooth shapes) coexist in equilibrium. This phenomenon has not yet been observed. Two-dimensional droplets of disodium cromoglycate (DSCG) have been the object of a thorough experimental study [Y.-K. Kim \emph{et al.},  J. Phys.: Condens. Matter \textbf{25}, 404202 (2013)]. We show that our theory is in good quantitative agreement with these data and extract from them what promises to be a more accurate estimate for the isotropic surface tension at the nematic/melt interface of DSCG.
\end{abstract}

\maketitle
\section{Introduction}\label{sec:intro}
Chromonic liquid crystals (CLC), also called \emph{chromonics}, form a very peculiar class of liquid crystals (LC).
These latter are systems constituted by basic \emph{mesogens} (either single molecules or molecular assemblies), which  
possess orientational ordering, and sometimes---under appropriate circumstances---also positional ordering.
Basically, there exist two classes of LCs: \emph{thermotropic} LCs, in which the ordering process is temperature-driven,
and \emph{lyotropic} LCs, in which the mesogenic units are molecular assemblies in a solvent, and concentration is the driving parameter for ordering. 
These latter are found in detergents and living organisms alike~\cite{ogolla:assembly}.

In particular, CLCs are lyotropics formed by certain dyes, drugs, and short nucleic-acid oligomers in aqueous solutions \cite{lydon:chromonic,dickinson:aggregate,tam-chang:chromonic,mariani:small,zanchetta:phase,nakata:end-to-end}
Since most biological processes take function normally in these types of solutions, 
it is no wonder that interest in CLCs has recently surged for possible applications in medical sciences. 
But this is not the only reason that makes them special (or rather unique). A number of informative, updated reviews are available on this topic \cite{lydon1998:chromonic,lydon:handbook,Lydon10,lydon:chromonic,dierking:novel}; they are  all  highly recommended. 

In ordinary lyotropic LCs, the aggregated molecules are bonded so strongly 
that neither size nor shape of the assemblies change appreciably upon varying moderately both temperature and concentration. 
Quite to the contrary, in CLCs, aggregation starts at very low concentrations.

CLC molecules are typically plank-shaped with aromatic cores and polar groups on their peripheries~\cite{ogolla:temperature}. They tend to stack face-to-face so as to form columns 
that order into a fluid nematic  phase or (for higher concentrations or lower temperatures) even in a solid-like medium phase,
where columns are organized parallel to one another with their centres arranged in an hexagonal pattern~\cite{lydon:chromonic}. 
The former phase is just the analogue of the nematic phase in ordinary thermotropics,
whilst the latter is the analogue of the columnar phase in discotic thermotropics.
It is the variability in size and shape of the supra-molecular columns that makes CLCs so unique.
There is also a wide variety 
of situations in the molecular organization: some are stacks of single molecules, others have more than a molecule 
in their cross-sections~\cite{ogolla:assembly}. These might seem to be minor details, but they may have momentous consequences at the macroscopic scale \cite{godinho:when}.

The aggregation process is often called \emph{isodesmic}, because the energy gain in adding a unit to a preexisting column 
(typically between $5$ and $10$ $kT$) does not depend on the length of the column. 
The isodesmic nature of the process results in a broad length column distribution, which is prone to the action of temperature. 
When the temperature is increased, the concentration of longer assemblies decreases. This is reflected by the elastic properties
of the phase, in a way that ordinary lyotropics do not exhibit~\cite{ogolla:temperature}. Further increasing the temperature results into a first order nematic-isotropic transition with a wide coexistence region ($5$-$10\,^\circ \mathrm{C}$). Conversely, when the temperature is decreased, short, disordered columns in the isotropic phase tend to grow and aggregate, eventually separating from the parent isotropic melt to form islands of ordered phase.

As customary, the nematic phase is described by the extra director field $\n$, which represents at the macroscopic level the local orientation of the supra-molecular rod-like aggregates. At the nematic/isotropic interface, an orientation-dependent surface tension arises that favours the (degenerate) tangential orientation of $\n$, in agreement with the purely entropic argument of Onsager~\cite{onsager:effects}, according to which a tangential anchoring would enhance the local translational freedom of columns at the interface.

We shall only be concerned with a two-dimensional problem, inspired by the experimental setting explored by Kim et al.~\cite{kim:morphogenesis}. There, CLC droplets in the nematic phase appeared surrounded by the isotropic phase,  their shape being the characteristic \emph{tactoids} (spindle-like) that were observed in aqueous dispersions of tobacco mosaic viruses since the seminal paper by Bernal and Fankuchen~\cite{bernal:x-ray}. Most reported droplets were \emph{bipolar}, with point defects of $\n$ at the pointed poles. 

In this paper, we adapt to the two-dimensional setting our theory for the representation of tactoids recently presented in \cite{paparini:nematic}. We shall achieve two main results. 

First, we predict a \emph{shape bistability}, which seems characteristic of the two-dimensional setting. We identify a range of droplet's areas where two distinct shapes could be observed, one tactoidal, as expected, and the other \emph{discoidal} (smooth). A sort of \emph{shape} coexistence thus parallels the phase coexistence observed in these materials. This regime manifests itself for droplets larger than those reported in \cite{kim:morphogenesis}; to our knowledge, it has not yet been observed.

Second, we use the very detailed data of \cite{kim:morphogenesis} to compare the observed shapes with those predicted by our theory. We extract an estimate for the isotropic component of the surface tension at the droplet's interface, which turns out to comparable in order of magnitude to the typical values measured for standard thermotropic materials ($\sim 10^{-5}\, \mathrm{J/m}^{2}$, see \cite[p.\,495]{kleman:soft}).

The paper is organized as follows. In Sec.~\ref{sec:retraction}, we recall from \cite{paparini:nematic} our theory and adapt it to the present two-dimensional setting. The optimal shapes of bipolar droplets that minimize the total free-energy functional are derived and discussed in Sec.~\ref{sec:optimal_shapes}, where  we illustrate in detail  the bistability scenario that we envision. We show, in particular, how the critical values of the droplet's area that delimit the corresponding shape hysteresis  depend on the elastic constants of the material. Section~\ref{sec:experiments_comparison} is devoted to the comparison with the experimental study \cite{kim:morphogenesis} that has motivated this work. We contrast our predicted shapes to the observed ones and, encouraged by their agreement, we estimate the isotropic component of the surface tension. Finally, in Sec.~\ref{sec:conclusion}, we summarize our conclusions and  comment on possible further extensions of our study. The paper is closed by two mathematical appendices, where we collect computational details  and auxiliary results needed in the main text, but inessential to its comprehension.

\section{Two-dimensional setting}\label{sec:retraction}
Here, we set our theoretical scene; we shall first recall the energetics of a CLC drop squeezed between two parallel plates and we shall then describe both its outer profile  and inner direction field. 

\subsection{Energetics}\label{sec:energetics}
In classical liquid crystal theory, the nematic director field $\n$ describes the average of the molecules that constitute the material. The elastic distortions of $\n$ are locally measured by its gradient $\nabla\n$, which may become singular where the director exhibits \emph{defects}, that is, discontinuities in the field  $\n$.  The bulk free energy is the following functional,
\begin{equation}
\label{eq:free_bulk}
\free_\mathrm{b}[\body,\n]:=\int_{\body} f_\mathrm{OF}(\n,\nabla\n)\dd V,
\end{equation}
where $\body$ is the region occupied by the material, $\dd V$ is the volume element, and $f_\mathrm{OF}$ is the Oseen-Frank free-energy density. Letting the latter be the most general frame-indifferent function, even  in $\n$ and quadratic in $\nabla\n$, one arrives at   (see, e.g., \cite[Ch.\,3]{virga:variational})
\begin{equation}\label{eq:frank_energy}
f_\mathrm{OF} := \frac{1}{2}K_{11}(\diver\n)^{2} + \frac{1}{2}K_{22}(\n\cdot\curl\n)^{2} + \frac{1}{2}K_{33}|\n\times\curl\n|^{2} + K_{24}\big(\tr(\nabla\n)^{2}-(\diver\n)^{2}\big),
\end{equation}
where $K_{11}$, $K_{22}$, $K_{33}$, and $K_{24}$ are the \emph{splay}, \emph{twist}, \emph{bend}, and \emph{saddle-splay} elastic constants, respectively. They are material moduli characteristic of each nematic  liquid crystal, corresponding to  four independent orientation fields, each igniting a single distinctive  term in \eqref{eq:frank_energy}.\footnote{Recently, an equivalent modal decomposition has been put forward for $f_\mathrm{OF}$ \cite{selinger:interpretation}, which has also been given a graphical representation in terms of an \emph{octupolar} tensor \cite{pedrini:liquid}.}

The energy $f_\mathrm{OF}$ is meant to penalize all distortions of $\n$ away from the unifom alignment (in whatever direction); to this end, the elastic constants in \eqref{eq:frank_energy} must satisfy Ericksen's inequalities \cite{ericksen:inequalities},
\begin{equation}
\label{eq:ericksen_inequalities}
K_{11}\geqq K_{24}\geqq0,\quad K_{22}\geqq K_{24}\geqq0,\quad K_{33}\geqq0.
\end{equation} 
CLCs in  three-dimensional space exhibit a different behavior: in cylindrical capillary tubes subject to degenerate tangential boundary conditions,\footnote{That is, when $\n$ is bound to be tangent to the boundary, but oriented in any direction.} the director $\n$ has been seen to steer away from the uniform alignment along the cylinder's axis \cite{jeong:chiral_2014,jeong:chiral_2015,davidson:chiral,nayani:spontaneous,javadi:cylindrical}. Two symmetric twisted configurations (left- and right-handed) have been observed in capillaries, each variant occurring with the same likelihood, as was to be expected from the lack of chirality in the molecular aggregates that constitute these materials.

Despite the clear indication that CLC's ground state differs from the uniform alignment presumed in $f_\mathrm{OF}$, the Oseen-Frank theory has been applied to rationalize the experiments with capillary tubes \cite{jeong:chiral_2014,jeong:chiral_2015,davidson:chiral,nayani:spontaneous}, at the cost of taking $K_{22}<K_{24}$, which violates one of Ericksen's inequalities \eqref{eq:ericksen_inequalities}. Since such a violation would make $f_\mathrm{OF}$ unbounded below, the legitimacy of these theoretical treatments is threatened by a number of mathematical issues that will be tackled elsewhere \cite{paparini:paradoxes}. Here, we can spare this issue, as our setting is two-dimensional. The region $\body$ will be a thin blob representing a droplet of CLC surrounded by its isotropic melt, squeezed between two parallel plates at the distance $h$ from one another.

\begin{figure}[h]
	\centering
	\includegraphics[width=0.4\linewidth]{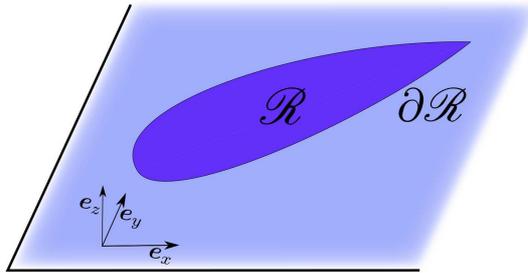}
	\caption{Two-dimensional domain $\region$ with area $A_0$ in the $(x,y)$ plane. It represents the cross-section of a drop squeezed between two parallel substrates $h$ apart and surrounded laterally by the isotropic phase. In accord with   the experimental observations in \cite{kim:morphogenesis}, both shape and director field are uniform across the gap of  thickness $h$ along $\e_z$.}
	\label{fig:shape_a}
\end{figure}

Formally, $\body=\region\times[-\frac{h}{2},\frac{h}{2}]$, where $\region$ is a region with piecewise smooth boundary $\partial\region$ in the $(x,y)$ plane of a Cartesian frame $\framec$ (see Fig.~\ref{fig:shape_a}).
The field $\n$ lies in the same plane as $\region$ and it is thought of as being extended uniformly through the blob's thickness. For a \emph{planar} field of this sort, the twist term in \eqref{eq:frank_energy} vanishes identically, as $\curl\n\parallel\e_z$, and so the twist constant $K_{22}$ will play no role here (and those Ericksen's inequalities that involve it can altogether be ignored). Thus, leaving aside the possibly controversial issue arising in general from the violation of one Ericksen's inequality, we remain assured that in the present setting the Oseen-Frank energy-density $f_\mathrm{OF}$ is bounded below and the free energy $\free_\mathrm{b}$ can be safely minimized.

We assume that a given mass of material constitute the blob $\body$, and so its volume $V_0$ is prescribed by the incompressibility constraint (which is satisfied to a large degree of approximation). Consequently, the area $A_0$ of $\region$ is prescribed, as $V_0=A_0h$. We further assume that the plates squeezing both $\body$ and the surrounding melt exert a degenerate tangential anchoring on the nematic director $\n$, so that, in light of the constraint on the area of $\region$, the additional anchoring energy can be treated as an inessential additive constant. 

This is not the case for the surface energy $\free_\mathrm{s}$ at the interface between $\body$ and the surrounding melt. Here we assume that this energy is represented by the Rapini-Papoular formula \cite{rapini:distortion}, 
\begin{equation}
\label{eq:free_sup}
\free_{\mathrm{s}}[\region,\n]=h\int_{\boundaryR}\gamma(1+\omega(\n\cdot\normal)^2) \dd A,
\end{equation}
where $\gamma>0$ is the \emph{isotropic} surface tension of the liquid crystal in contact with its melt, $\omega$ is a dimensionless \emph{anchoring strength}, which we take to satisfy $\omega\geqq0$, $\normal$ is the outer unit normal to $\boundaryR$, and $\dd A$ is the area element. 
Thus $\free_{\mathrm{s}}$ is minimized when $\n$ lies tangent to $\boundaryR$. Henceforth, we shall enforce this minimum requirement as a constraint on $\n$,
\begin{equation}
\label{eq:degenerate_boundary_condition}
\n\cdot\normal\equiv0, \quad \hbox{on} \quad \boundaryR,
\end{equation}
save checking ultimately this assumption with appropriate energy comparisons (see Appendix~\ref{sec:alpha_safeguard}) which ensures that such a \emph{tangential anchoring} is not broken. In short, the validity of \eqref{eq:degenerate_boundary_condition} requires that the droplet is not too \emph{small} in a sense that will be made precise below.

To sum up, the total free energy $\free:=\free_{\mathrm{b}}+\free_{\mathrm{s}}$ of the system is given by the functional
\begin{equation}
\label{eq:free_energy_functional}
\free[\region,\n]:=h\left(\int_{\region} f_\mathrm{OF}\dd A+\gamma \ell(\boundaryR)\right),
\end{equation}
subject to \eqref{eq:degenerate_boundary_condition} and to the isoperimetric constraint
\begin{equation}
\label{eq:area_constraintA0}
A(\region)=A_0,
\end{equation}
where  $A$ and $\ell$ are the area and length measures, respectively.

\subsection{Admissible Shapes}\label{sec:admissible_shapes}
\begin{figure}[h]
	\centering
	\includegraphics[width=0.4\linewidth]{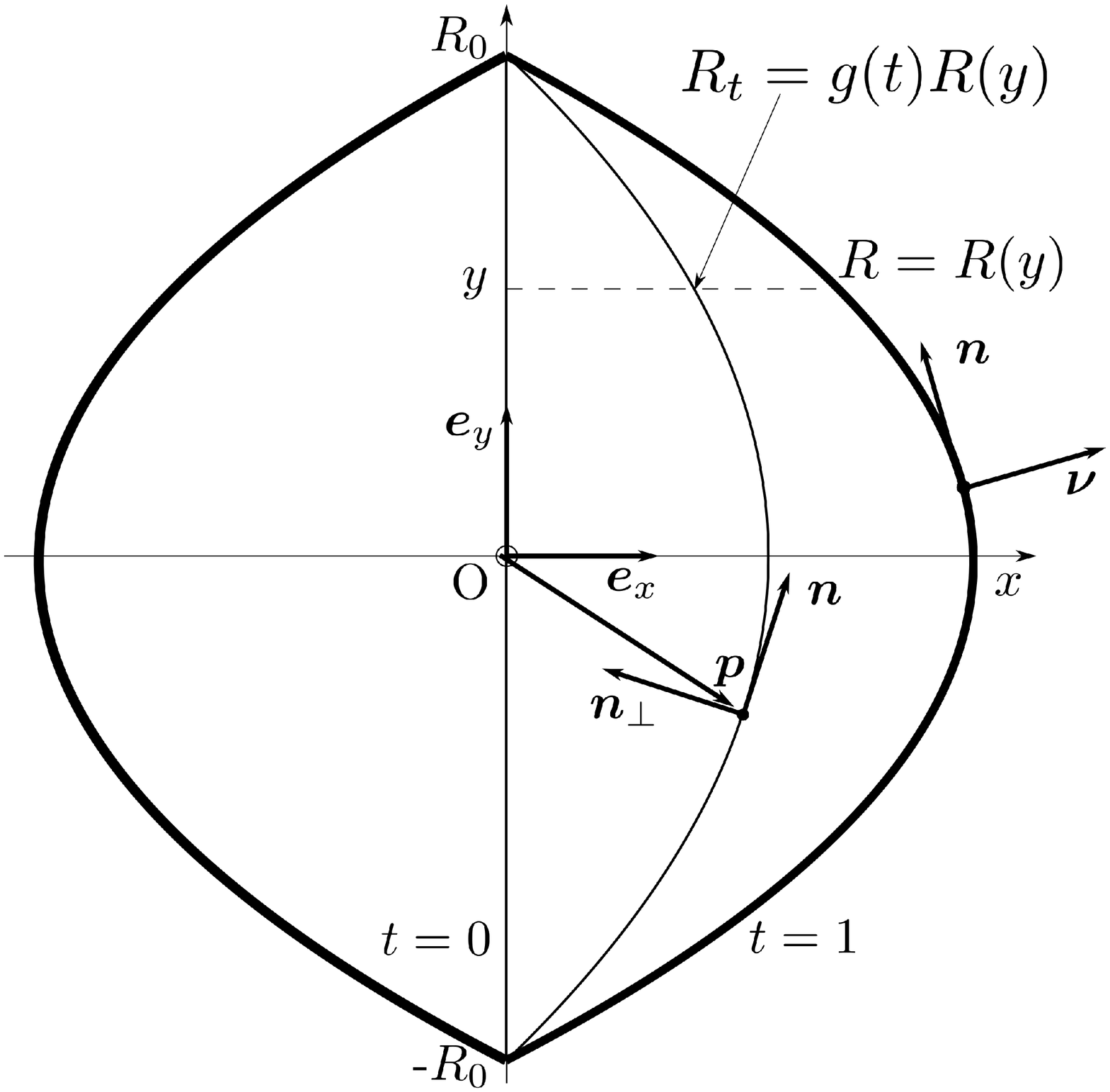}
	\caption{The function $R(y)$ and its retraction $R_t(y)=g(t)R(y)$ represent (half) the boundary $\boundaryR$ and retracted inner curves $\boundaryR_t$ for generic  $t\in[0,1]$. The nematic director $\n$ is the unit vector field everywhere tangent to the retracted curves; $\nper=\e_z\times\n$ is the orthogonal field, and $\normal$ is the outer unit normal to $\boundaryR$, where $\normal=-\nper$. The function $R$ represents half a drop; the other half is obtained by mirror symmetry about the $y$ axis.}
	\label{fig:shape_b}
\end{figure}

No preferred direction is present on the substrates that squeeze the drop, and so, contemplating tactoids among the possible equilibrium shapes of $\region$ (as well as other smooth shapes), we assume that these are mirror symmetric about two orthogonal axes, one joining the possible sharp tips of the boundary $\boundaryR$. We denote by $y$ the latter axis and by $x$ the orthogonal symmetry axis.\footnote{Clearly, due to the absence of a privileged orientation on the bounding substrates, the $y$ axis could indeed be oriented in any direction. If several drops are present, their axes would be isotropically distributed.}  Thus, only  half of the curve that bounds $\region$ needs to be described, the other half being obtained by mirror symmetry. We take this curve to be represented as $x=R(y)$, where $y$ ranges in the interval $[-R_0,R_0]$, with $R_0$  to be determined, and $R$ is a smooth, even function such that 
\begin{equation}
	\label{eq:R_definition}
	R(R_0)=0\quad\text{and}\quad R'(0)=0,
\end{equation}
where a prime $'$ denotes differentiation with respect to $y$ (see Fig.~\ref{fig:shape_b}).
The points where $R$ vanishes correspond to the \emph{poles} of $\region$. Whenever $R'(R_0)$ is finite, $R$ represents a \emph{tactoid}, as the outer unit normal $\normal$ to $\boundaryR$ is discontinuous at the poles. Smooth shapes correspond to $R'(R_0)=-\infty$. 

We call $\Req$ the radius of the \emph{equivalent} disc, which has area $A_0$, and we denote by $\mu$ the dimensionless length of the semi-axis of the drop,
\begin{equation}
	\label{eq:mu_definition}
	\mu:=\frac{R_0}{\Req}.
\end{equation}
Hereafter, we shall rescale all lengths to $\Req$ (while keeping their names unchanged, to avoid typographical clutter). With this normalization, 
the area constraint \eqref{eq:area_constraintA0} reads simply as
\begin{equation}
	\label{eq:area_rescaled}
	\int_{-\mu}^{\mu} R(y) \dd z=\frac\pi2.
\end{equation}

\subsection{Director Retraction}\label{sec:director_retraction}
The shape of $\region$ is unknown and needs to be determined. Since $\n$ is tangent to $\boundaryR$, following \cite{paparini:nematic} we device a method that also derives $\n$ inside $\region$ from the knowledge of $\boundaryR$, thus reducing the total free energy $\free$ to a pure \emph{shape} functional. This is achieved by \emph{retracting} $\boundaryR$ inside $\region$ with its tangent field $\n$.

Formally, we define a function $R_t(y)=g(t)R(y)$, where $t\in[0,1]$ and $g$ is an increasing monotonic function such that $g(0)=0$ and $g(1)=1$. The graph of $R_t$, shown in Fig.~\ref{fig:shape_b}, represents the \emph{retraction} of $\boundaryR$ that borders an inner domain $\region_t\subseteq\region$. All domains $\region_t$ are nested one inside the other as $t$ decreases towards $0$. For $t=0$, $\region_t$ reduces to the $y$ axis. The advantage of this method is that it also affords to describe the inner director $\n$ as the field everywhere tangent to the family of curves $\boundaryR_t$. All director fields obtained by this geometric construction are \emph{bipolar}, in that they have two point defects at the poles; in the language of Mermin~\cite{mermin:topological}, they are \emph{boojums} with topological charge $m=+1$ (see also \cite[p.\,501]{kleman:soft}).

It is shown in Appendix~\ref{sec:retracted_field} how to compute the area element $\dd A$ in the $(t,y)$ coordinates and how to express $\nabla\n$ in the orthonormal frame $(\n,\nper)$, where $\nper:=\e_z\times\n$, in terms of the functions $R(y)$ and $g(t)$.

In the rescaled variables  $y$ and $R(y)$,  an appropriate dimensionless form of $\free$ in \eqref{eq:free_energy_functional} is then given by
\begin{align}
\label{eq:F_energy_functional} 
F[\mu;R]:=\frac{\free[\body]}{K_{11}h}=\int_{-\mu}^{\mu}&\left\{\left[\frac{R'}{R}-\dfrac{R''}{R'}+\frac{1}{8}\dfrac{RR''^2}{R'^3}\left(3+k_3\right)\right]\arctan R'+\frac{R''}{1+R'^2}+\right. \nonumber\\
&\left.+\frac{1}{8}\frac{RR''^2}{(1+R'^2)^2}\left[(k_3-5)-\frac{1}{R'^2}(k_3+3)\right]\right\}+2\alpha\sqrt{1+R'^2}\dd y,
\end{align}
where 
\begin{equation}
\label{eq:elastic_constants_rescaled}
k_3:=\frac{K_{33}}{K_{11}}
\end{equation}
is a \emph{reduced bend} constant and
\begin{equation}
\label{eq:alpha}
\alpha:=\frac{\gamma\Req}{K_{11}}
\end{equation}
is a \emph{reduced area}. Equivalently, $\alpha=\Req/\xi_\mathrm{e}$, where $\xi_\mathrm{e}$ is the de Gennes-Kleman \emph{extrapolation length} \cite[p.\,159]{kleman:soft}.\footnote{Thus, when we say that a drop is either small or large, we mean precisely that either $\alpha\ll1$ or $\alpha\gg1$, respectively.\label{foot:alpha}} 

\begin{figure}[h] 
	\includegraphics[width=.2\linewidth]{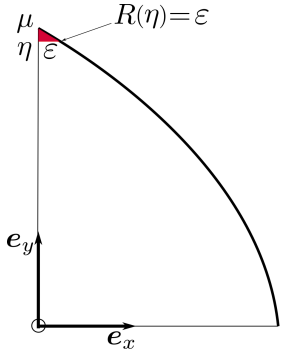}
	\caption{Isotropic defect core (in red) at the tip of a drop (only a quarter is shown). In our dimensionless  units (rescaled to $\Req$), $\eta$ is defined by \eqref{eq:y_bar}. A sensible value for the parameter $\vae$, which here is out of scale, is $\vae\approx10^{-3}$. A similar construction also applies to a drop with smooth poles (not shown here).}
	\label{fig:cut}
\end{figure}

The  reduced functional $F$ in \eqref{eq:F_energy_functional} suffers from a typical pathology of two-dimensional director theory: it diverges logarithmically to $+\infty$ near defects. Here, the culprit is the integrand $\frac{R'}{R}\arctan R'$, which is not integrable at $y=\pm\mu$.
Following a well established practice (see, e.g.,\cite[p. 171]{degennes:physics}), we imagine that the energy concentration near defects causes a localized transition to the isotropic phase, which constitutes a \emph{defect core} (whose fine structure is better explored within Ericksen's theory  \cite{ericksen:liquid}). The energy associated with such a phase transition is proportional to the core's area and will be taken as approximately fixed. Moreover, for simplicity, instead of considering a circular core,  which in the most common choice, we take it in the shape of the tapering drop's tip. Letting $r_\mathrm{c}$ denote the core's size, we set $r_\mathrm{c}=\vae\Req$ and restrict $y$ to the interval $[-\eta,+\eta]$, where $\eta$ is defined by 
\begin{equation}
\label{eq:y_bar}
R(\eta) = R (-\eta) = \varepsilon,
\end{equation}
and so depends indirectly on $\vae$ (see Fig.~\ref{fig:cut}).
For $\Req$ of the order of $10\mu\mathrm{m}$, it is reasonable to take $\vae\approx10^{-3}$, as we shall do here, which corresponds to $r_\mathrm{c}$ of the order of $10\mathrm{nm}$. The integral in \eqref{eq:F_energy_functional} will hereafter be limited to the interval $[-\eta,+\eta]$, so that it will always converge. The extra energy stored in the defects, being approximately constant, will play no role in our quest for the equilibrium shape of squeezed drops.

\subsection{Special Family of Shapes}\label{sec:degenerate_substrates_family_shape}
Here, we follow closely \cite{paparini:nematic}, albeit in a two-dimensional setting, in an attempt to restrict the admissible shapes of drops to a special family amenable to a simple analytical treatment. The
admissible drop profiles will  be described by the function\footnote{For $b=0$, $R$ in \eqref{eq:profile} reduces to he parabolic profile considered in \cite{williams:nematic}.}  
\begin{equation}
\label{eq:profile}
R(y)=a(\mu^2-y^2)+b\sqrt{\mu^2-y^2},
\end{equation}
where $a$ and $b$ are real parameters that must be chosen subject to the requirements that $R(y)\geqq0$ for all $-\mu\leqq y\leqq\mu$  and that \eqref{eq:area_rescaled} is satisfied. It is a simple matter to show (see also \cite{paparini:nematic}) that $a$ and $b$ can be expressed in terms of the free  parameters $(\phi,\mu)$ that span the configuration space $\conf:=\{(\phi,\mu): 0\leqq\phi\leqq\frac{3\pi}{4},\ \mu>0\}$. Precisely,
\begin{equation}
\label{eq:a_b_representation}
a=\frac{1}{\mu^{3}}\frac{\pi\cos\phi}{h(\phi)},\quad b=\frac{1}{\mu^{2}}\frac{\pi\sin\phi}{h(\phi)}\quad\text{with}\quad h(\phi):=\frac{8}{3}\cos\phi+\pi \sin\phi>0,\quad 0\leqq\phi\leqq\frac{3\pi}{4}.
\end{equation}

\begin{figure}[h] 
	\includegraphics[width=.6\linewidth]{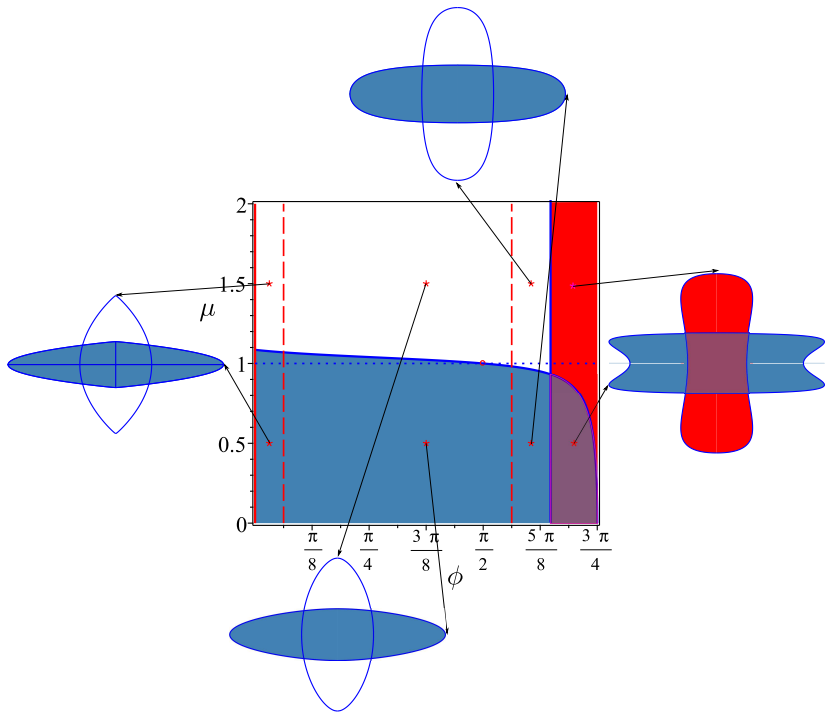}
	\caption{Configuration space with the admissible shapes described by \eqref{eq:profile}. The blue region below the graph of the function $\varpi(\phi)$ in \eqref{eq:prolate_curve} represents all prolate shapes. The red  strip for $\phi_\mathrm{c}\leqq\phi\leqq\frac{3\pi}{4}$ represents the concave shapes that we call \emph{butterflies}; all shapes falling on the left of this strip are convex. The circular disc is represented by the point $(\frac\pi2,1)$, marked by a red circle. We also call \emph{tactoids} the shapes for $0\leqq\phi\leqq\frac{\pi}{16}$ (\emph{genuine} tactoids, only those for $\phi=0$, marked by a red line), \emph{discoids} those for $\frac{\pi}{16}\leqq\phi\leqq\frac{9\pi}{16}$, and \emph{b\^{a}tonnet} those for $\frac{9\pi}{16}\leqq\phi\leqq\phi_\mathrm{c}$, see also Table~\ref{tab:taxonomy}, and  Fig.~\ref{fig:gallery} for a fuller gallery of shapes. The barriers marking transitions from one family of shapes to another are represented by vertical dashed lines.}
	\label{fig:configuration_space}
\end{figure}

Shapes with different qualitative features correspond to different regions of $\conf$, as illustrated in Fig.~\ref{fig:configuration_space}. \emph{Prolate} shapes are characterized by
\begin{equation}
\label{eq:prolate_curve}
\mu\geqq\varpi(\phi):=\sqrt{\frac{\pi(\cos\phi+\sin\phi)}{h(\phi)}},
\end{equation}  
whereas \emph{oblate} shapes are characterized by $\mu<\varpi(\phi)$. Moreover, shapes represented by \eqref{eq:a_b_representation} are convex for $0\leqq\phi\leqq\phi_\mathrm{c}:=\arccot\left(-\frac12\right)\doteq2.03$ and concave for $\phi_\mathrm{c}<\phi\leqq\frac{3\pi}{4}$. The latter are represented by the red strip in Fig.~\ref{fig:configuration_space};
we call them  \emph{butterflies}: their waist narrows as $\phi$ approaches the boundary of $\mathsf{S}$ at $\phi=\frac{3\pi}{4}$, where it vanishes altogether and the droplet splits in two.

$\region$ has pointed tips only whenever $R'(\mu)$ is finite;  according to \eqref{eq:a_b_representation}, the only value of $\phi$ that makes $b$ vanish is $\phi=0$. We call \emph{genuine} these tactoids. For small enough values of $\phi$ the shape represented by \eqref{eq:profile} via \eqref{eq:a_b_representation} cannot be visually distinguished from pointed tactoids; we find the conventional barrier at $\phi=\frac{\pi}{16}$ appropriate to delimit the realm of tactoids (genuine or not). Further increasing $\phi$, $\region$ has tips that look smoother, justifying our calling them \emph{discoids}. A conventional barrier is set  at $\phi=\frac{9\pi}{16}$ to mark  where discoids evolve into  little batons, for which we use the French word \emph{b\^atonnet}. The  names used to identify different shapes of $\region$ and the corresponding strips in $\conf$ where they are found are recalled in  Table~\ref{tab:taxonomy}. 
\begin{table}[h]
	\caption{We identify four strips in configuration space $\mathsf{S}$, which correspond to four qualitatively different shapes for a region $\region$ represented by \eqref{eq:profile} via \eqref{eq:a_b_representation}. Here $\phi_\mathrm{c}=\arccot\left(-\frac12\right)\doteq2.03$.}
\begin{ruledtabular}
	\begin{tabular}{ccccc}
		Tactoids&Discoids&B\^{a}tonnet&Butterflies\\
		\hline\\
		$0\leqq\phi<\dfrac{\pi}{16}$&$\dfrac{\pi}{16}<\phi<\dfrac{9\pi}{16}$&$\dfrac{9\pi}{16}<\phi<\phi_\mathrm{c}$&$\phi_\mathrm{c}<\phi\leqq\dfrac{3\pi}{4}$\\
		\\
	\end{tabular}
\end{ruledtabular}
\label{tab:taxonomy}
\end{table}

\begin{figure}[h]
	\centering
	\begin{subfigure}[c]{0.19\linewidth}
		\centering
		\includegraphics[width=\linewidth]{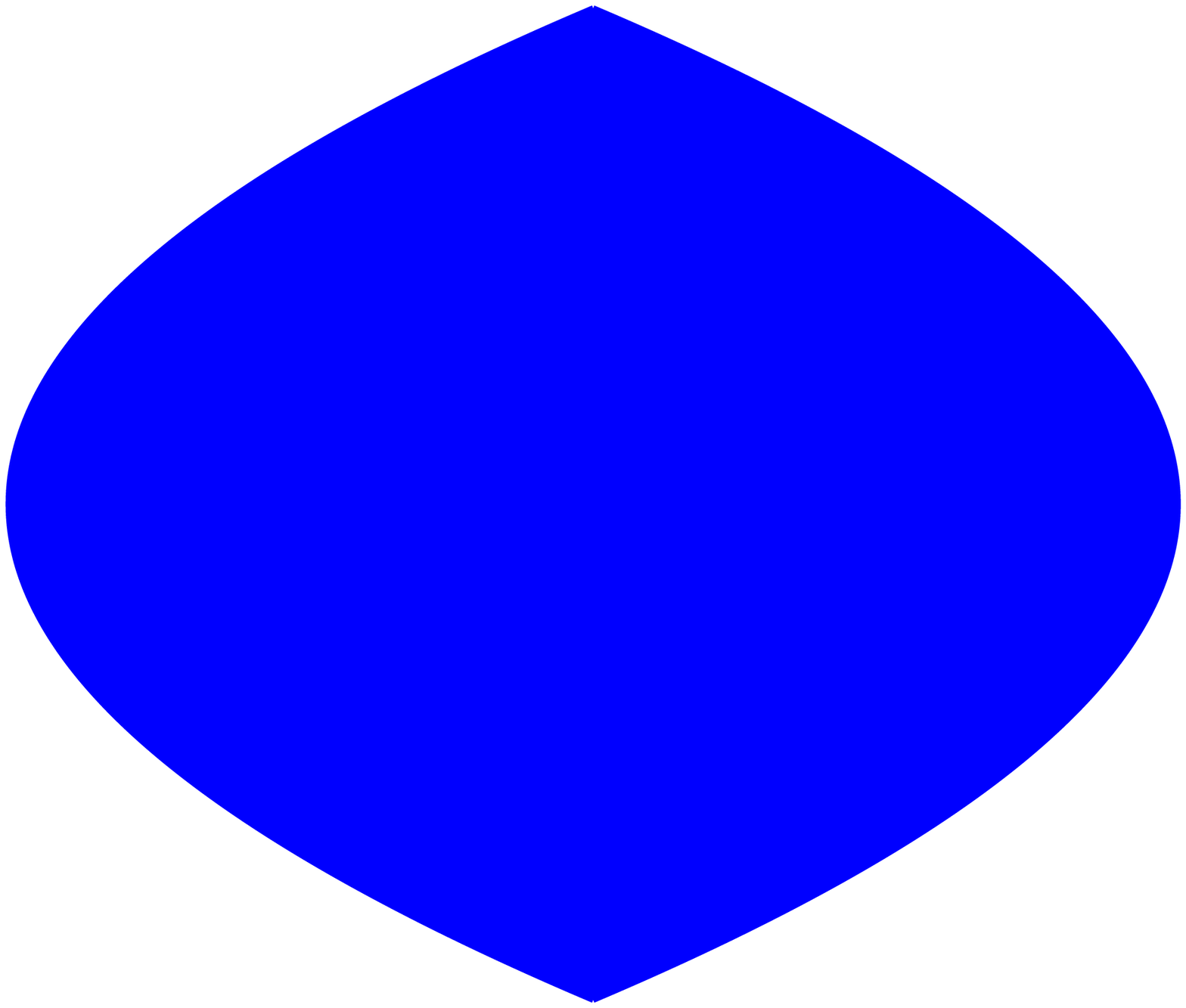}
		\caption{$\phi=0$\\genuine tactoid}
		\label{fig:gallery_0Pio16mu1}
	\end{subfigure}
	\begin{subfigure}[c]{0.19\linewidth}
	\centering
	\includegraphics[width=\linewidth]{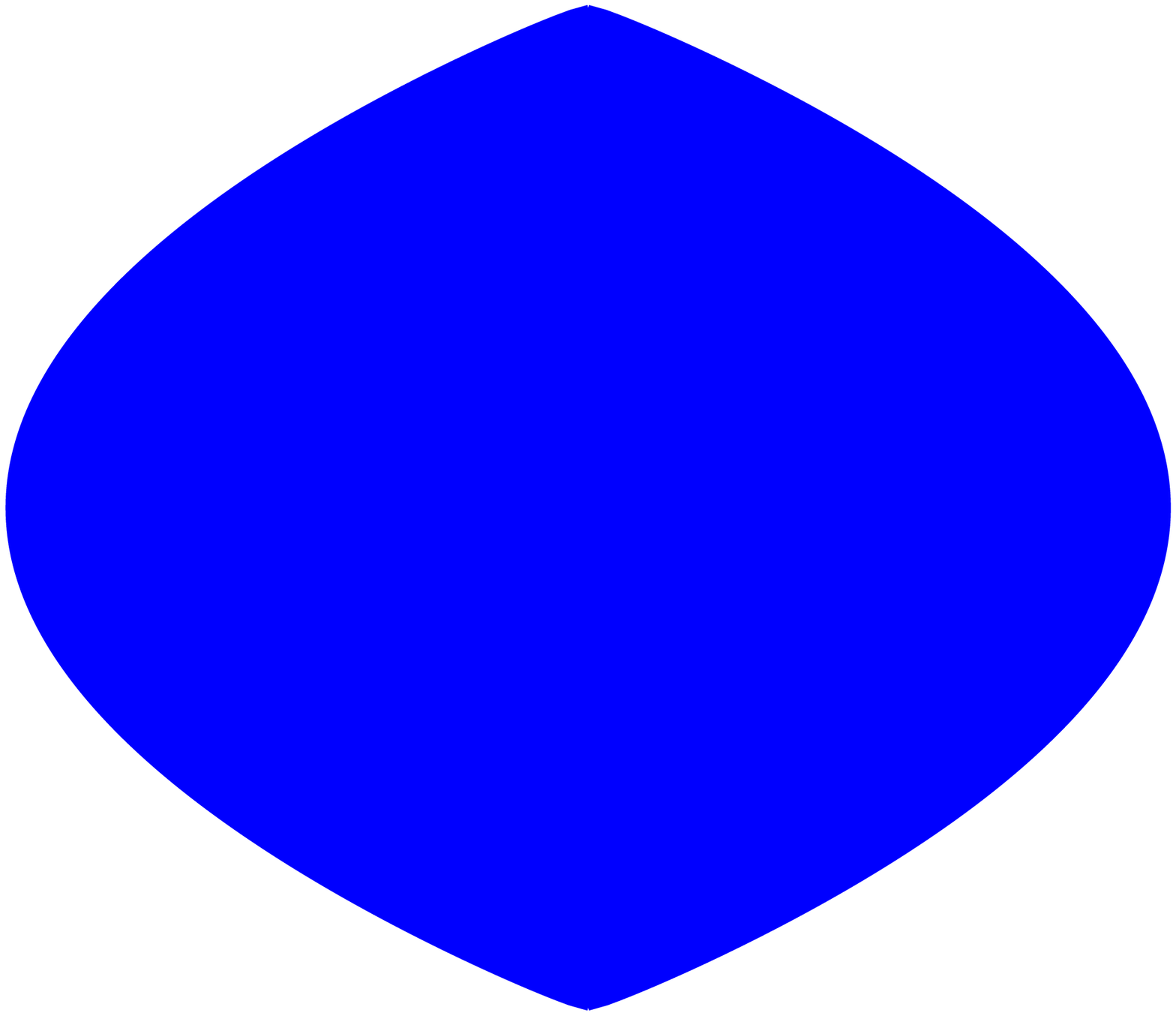}
	\caption{$\phi=\frac{\pi}{32}$\\tactoid}
	\label{fig:gallery_Pio32mu1}
\end{subfigure}
	\begin{subfigure}[c]{0.19\linewidth}
		\centering
		\includegraphics[width=\linewidth]{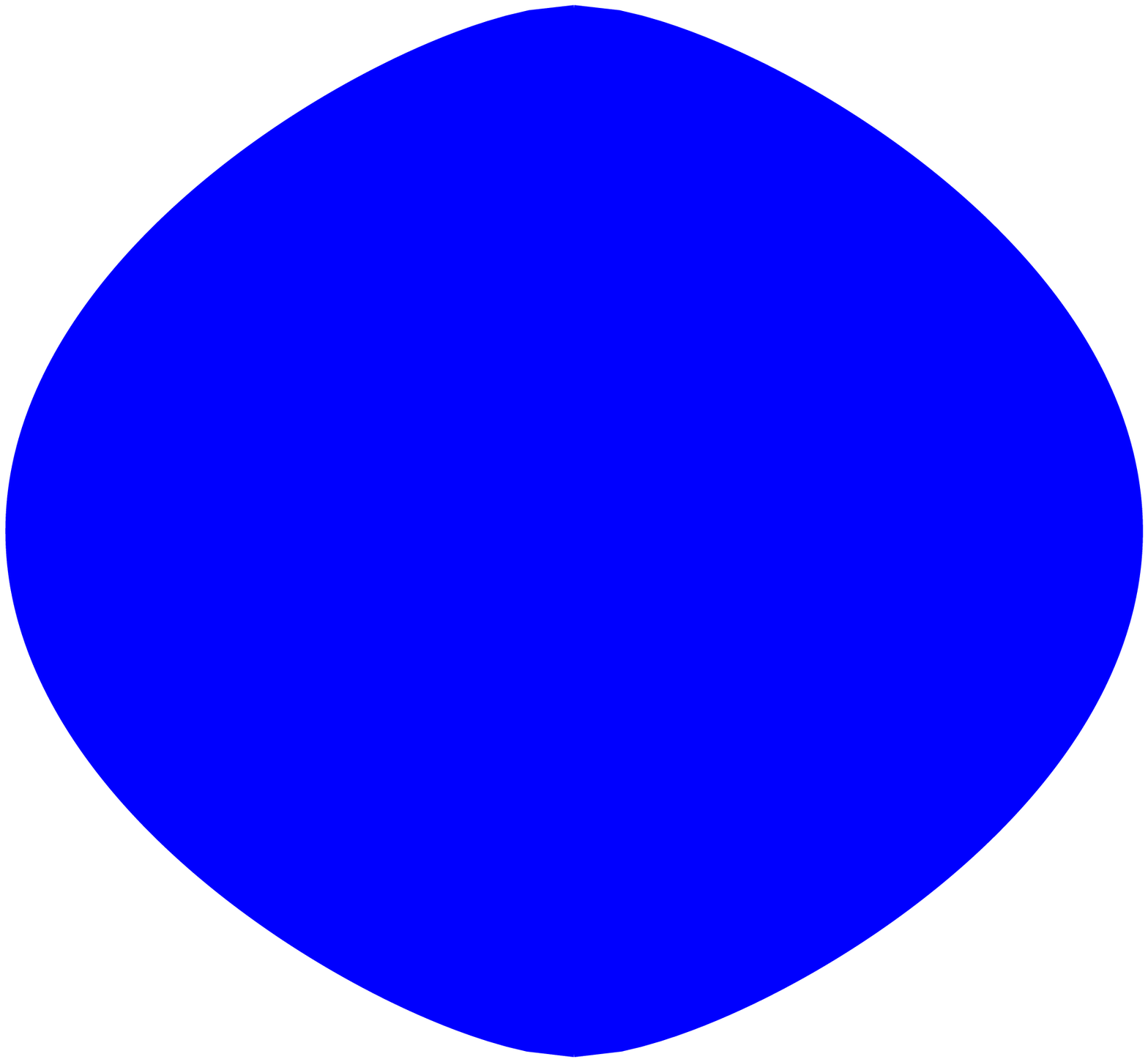}
		\caption{$\phi=\frac{\pi}{4}$\\discoid}
		\label{fig:gallery_4Pio16mu1}
	\end{subfigure}
	\begin{subfigure}[c]{0.19\linewidth}
	\centering
	\includegraphics[width=\linewidth]{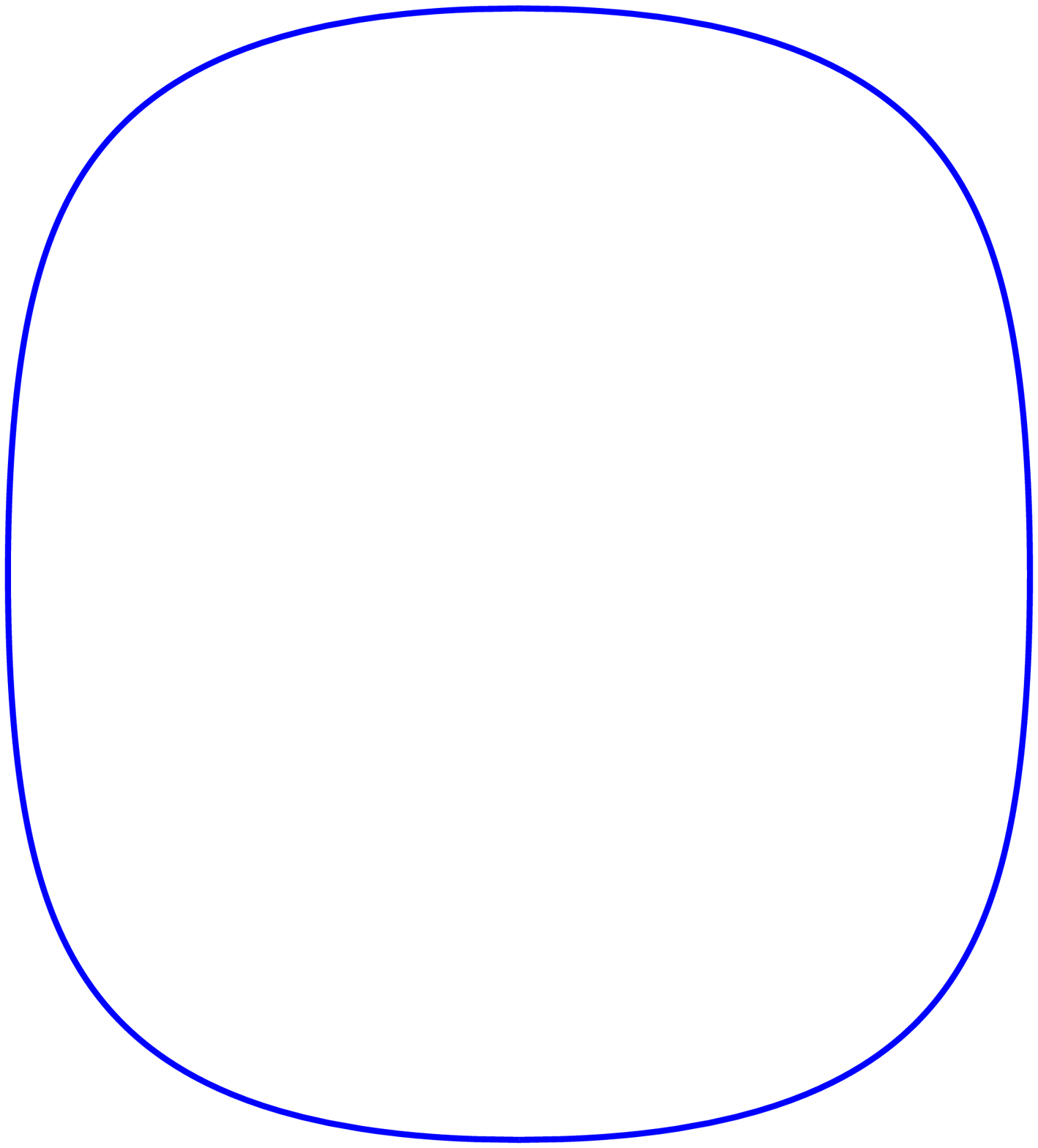}
	\caption{$\phi=\frac{10\pi}{16}$\\b\^{a}tonnet}
	\label{fig:gallery_10Pio16mu1}
\end{subfigure}
	\begin{subfigure}[c]{0.19\linewidth}
	\centering
	\includegraphics[width=\linewidth]{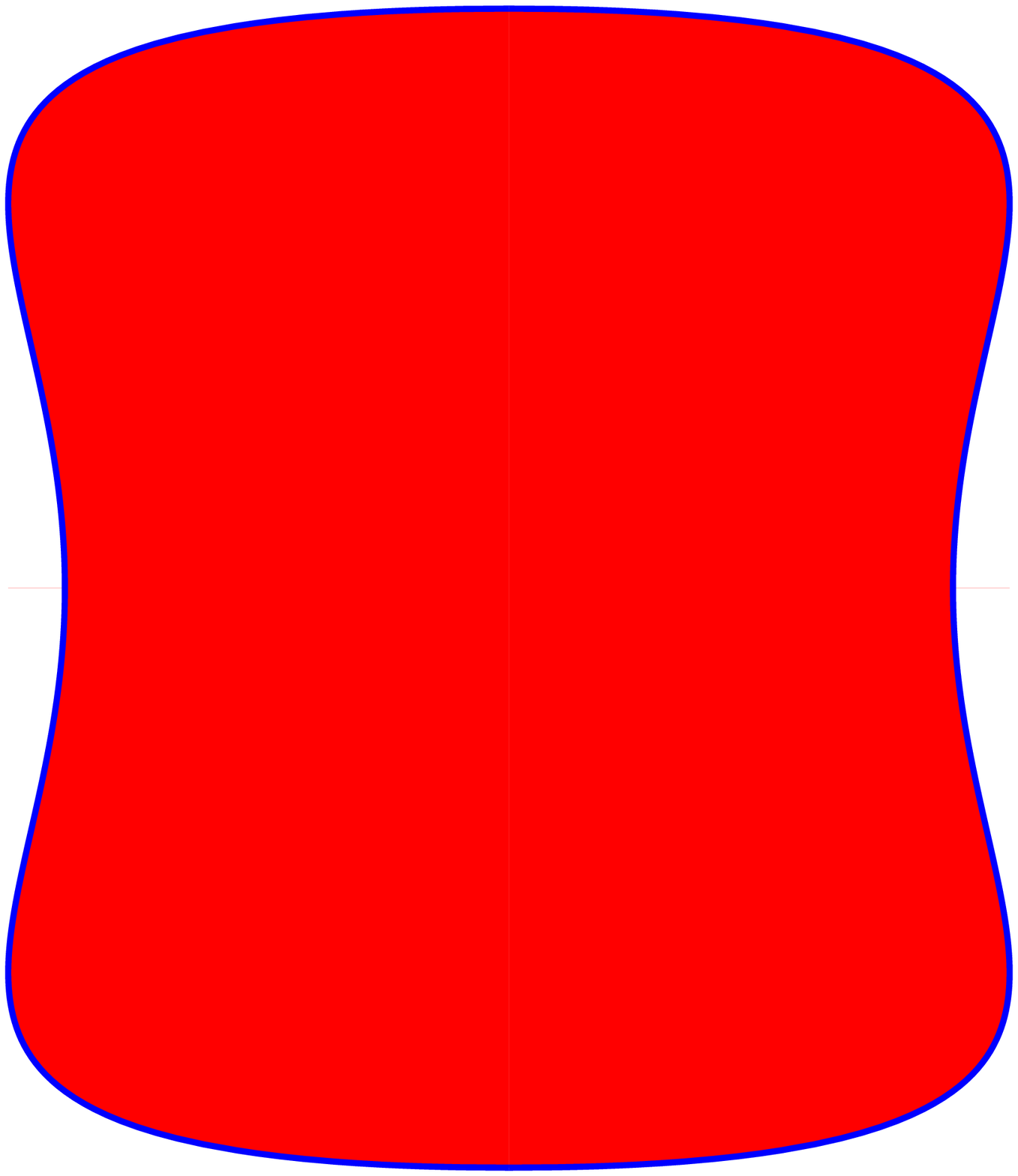}
	\caption{$\phi=\frac{11\pi}{16}$\\butterfly}
	\label{fig:gallery_11Pio16mu1}
\end{subfigure}
	\caption{Gallery of shapes illustrating for $\mu=1$ the  taxonomy introduced in Table~\ref{tab:taxonomy}.
		The color coding of the shapes is the same used in Fig.~\ref{fig:configuration_space}.}
	\label{fig:gallery}
\end{figure}

Fig.~\ref{fig:gallery} illustrates a gallery of shapes for $\region$ obtained from \eqref{eq:profile} and \eqref{eq:a_b_representation} for $\mu=1$  and a number of values of $\phi$ falling in the different types listed in Table~\ref{tab:taxonomy}. 

\section{Optimal Shapes}\label{sec:optimal_shapes}
This section is devoted to the study of the minimizers of the free-energy functional in \eqref{eq:F_energy_functional}. Our major result will be the prediction of a shape bistability, which has not yet been observed in this setting. Before describing this phenomenon, however, we need to make sure that the droplets we consider are not too small for the tangential anchoring condition in \eqref{eq:degenerate_boundary_condition} to be valid.

\subsection{Admissible Drop Sizes}\label{sec:degenerate_substrates_admissible volumes}
It is known \cite{virga:drops} that drops sufficiently small in three-dimensional space tend to break the director's tangential anchoring on their boundary, favouring the uniform alignment of $\n$ in their bulk. A simple heuristic argument tells us that a similar breaking would also take place in the present two-dimensional setting.  

The elastic cost of the bulk deformation scales as $Kh$, where $K$ is a typical elastic constant, while the surface anchoring energy scales as $\gamma\omega h\Req$, and so $\alpha$ \eqref{eq:alpha} estimates the ratio of the latter to the former. Thus, when $\alpha$  is sufficiently small, the bulk energy becomes dominant and it is minimized by the uniform alignment of $\n$, which breaks the tangential anchoring, undermining our analysis. In Appendix~\ref{sec:alpha_safeguard}, we perform an energy comparison that provides an estimate for the \emph{safeguard} vale $\alpha_\mathrm{s}$ of $\alpha$, above  which  tangential anchoring is expected to remain unbroken. We obtained the following explicit formula for $\alpha_\mathrm{s}$,
\begin{equation}
	\label{eq:alpha_safeguard}
\alpha_\mathrm{s}:=\frac{\frac{\pi}{4}\left(k_3-1-k_3\ln2-\ln\varepsilon\right)+\varepsilon}{j(\omega)-\left(\frac{\pi}{2}-\varepsilon\right)-\frac{\omega}{2} \pi \varepsilon},
\end{equation}
where $j(\omega)$ is the function defined in \eqref{eq:free_a_hom_degenerate}. For $\omega=5$, which is a choice supported by some evidence,\footnote{See, for example, \cite{puech:nematic} and \cite{kim:morphogenesis}.} $\alpha_\mathrm{s}\approx0.2k_3-0.5(1+\ln\vae)$. In particular,
for $k_3=1$ and $\varepsilon=10^{-3}$,   $\alpha_s\approx 3$, which will be our reference choice henceforth. Taking $K\sim1$-$10\,\mathrm{pN}$ as typical value for all elastic constants\footnote{This estimate is supported for example by \cite{zhou:elasticity_2014} for material such as DSCG, SSY, and PBG.} and $\gamma\sim10^{-5}\,\mathrm{J/m}^{2}$ as typical value for the surface tension of a chromonic liquid crystal in contact with its melt,\footnote{Discordant estimates of $\gamma$ have been given in the literature \cite{kim:morphogenesis,mushenheim2014:using,tortora:chiral}. Here we take the average order of magnitude found in these works (see also Sec.~\ref{sec:experiments_comparison} below for an independent justification of this choice).} by \eqref{eq:alpha} taking $\alpha>3$ means taking $\Req\gtrsim 0.3$-$3\, \mu\mathrm{m}$, which provides a lower bound on the admissible size of the droplets that can be treated within our theory.

\subsection{Shape Bistability}\label{sec:degenerate_substrates_minimizing_trajectories}
 
Finding analytically the minima of the \emph{reduced} free energy   $\Fa(\phi,\mu)$, the function defined on the configuration space $\conf$ by computing the functional $F\left[\mu;R\right]$ in \eqref{eq:F_energy_functional} on the special family of shapes in \eqref{eq:profile}, is simply impracticable. 
Thus, for a given choice of the elastic parameter $k_3$,  we evaluated numerically $\Fa$ and we sought its minimizers in $\conf$ for increasing $\alpha>\alpha_\mathrm{s}$.

We found out that there are two critical values of $\alpha$, $\alpha_1$ and $\alpha_2>\alpha_1$, such that for either $\alpha<\alpha_1$ or $\alpha>\alpha_2$, $F_\alpha$ attains a single (absolute) minimum in $\conf$, whereas it attains two (relative) minima for $\alpha_1\leqq\alpha\leqq\alpha_2$. There is a third critical value, $\alpha_\mathrm{b}\in(\alpha_1,\alpha_2)$, such that for  $\alpha=\alpha_\mathrm{b}$ the two minima of $F_\alpha$ are equal and its absolute minimizer abruptly shifts from one point in $\conf$ to another. Were the points of $\conf$ to represent  the different phases of a condensed system, this scenario would be described as a (first-order) \emph{phase transition}. In our setting, it more simply describes the (local) stability of two equilibrium shapes for a squeezed drop: for $\alpha_1<\alpha<\alpha_2$, both a tactoid and a discoid are local energy minimzers, the global minimum shifting from the former to the latter at $\alpha=\alpha_\mathrm{b}$. For $\alpha<\alpha_1$, the only equilibrium shape is a tactoid, whereas it is a discoid for $\alpha>\alpha_2$. The \emph{shape bistability} exhibited by this two-dimensional system will now be documented in more details.

\begin{figure}
	\centering
	\includegraphics[width=.5\linewidth]{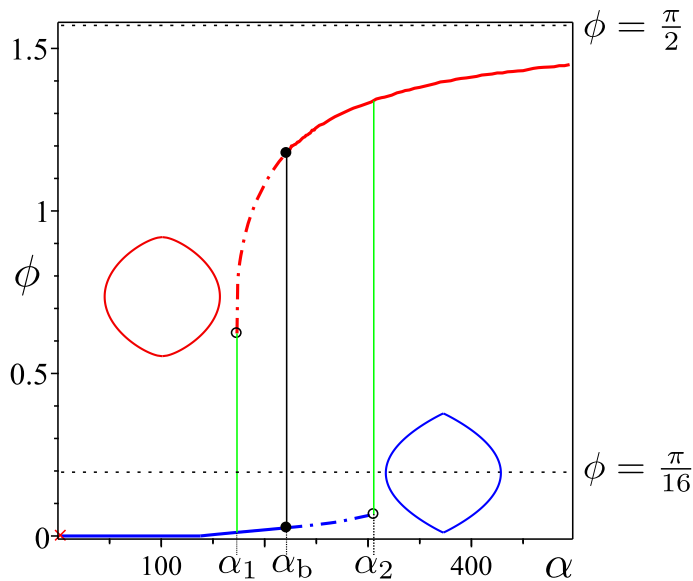}
	\caption{Bifurcation diagram for the parameter $\phi$ that describes the equilibrium profile \eqref{eq:profile} as a function of the bifurcation parameter $\alpha$. For every $k_3$, there are three critical values of $\alpha$, namely, $\alpha_1$, $\alpha_\mathrm{b}$, and $\alpha_2$. Solid lines represent global minima, while broken lines represent local minima.  Open circles mark the appearance or disappearance of a local minimum at $\alpha=\alpha_1$ and $\alpha=\alpha_2$. Two green lines delimit the coexistence interval $[\alpha_1,\alpha_2]$. At $\alpha=\alpha_\mathrm{b}$, both local minima are also global minima; two black dots mark the jump of the global minimum from one branch to the other. Here $k_3=1$, $\alpha_1=172$, $\alpha_\mathrm{b}=221$, and $\alpha_2=305$. The droplet's equilibrium shape is drawn for the critical values $\alpha_1$ and $\alpha_2$. }
	\label{fig:minimum_life_degeneratek31}
\end{figure}

We start by representing the equilibrium landscape in the language of bifurcation theory. Taking $\alpha$ as bifurcation parameter and $\phi$ as equilibrium shape representative, in
Fig.~\ref{fig:minimum_life_degeneratek31}
we illustrate the minima of  $\Fa$ for $k_3=1$: one falls in  $0<\phi<\frac{\pi}{16}$ (blue line), and so it is a tactoid, while the other falls in $\frac{\pi}{16}<\phi<\frac{\pi}{2}$ (red line) and is a discoid. Solid lines represent global minima, while broken lines represent local minima. Two separate local minima are also global minima for $\alpha=\alpha_\mathrm{b}$, where a perfect bistability is established between the two equilibrium branches. The tacoidal branch can be further continued,  as it is locally stable, until $\alpha$ reaches the critical value $\alpha_2$, where it ceases to exist altogether. Similarly,  as soon as $\alpha$ exceeds $\alpha_1$, the discoidal branch comes first into life as a locally stable equilibrium, which then becomes globally stable for $\alpha>\alpha_\mathrm{b}$. For $\alpha\in[\alpha_1,\alpha_2]$, tactoids and discoids coexist as optimal  shapes; both are metastable, one or the other is globally stable, according to whether $\alpha<\alpha_\mathrm{b}$ or $\alpha>\alpha_\mathrm{b}$. 

The dimensionless parameter $\alpha$ is ultimately related through \eqref{eq:alpha} to the amount of material trapped in the drop. So the coexistence interval $[\alpha_1,\alpha_2]$ corresponds to a window of areas $A_0$ for which two different shapes could be observed, more likely (and frequently) the one corresponding to the global minimum. If, ideally, one could gently pump material into a tactoidal drop, so as to follow the blue branch in Fig.~\ref{fig:minimum_life_degeneratek31}, the drop would continue to grow as a tactoid until the critical volume corresponds to $\alpha_2$, where a dynamical instability would presumably prompt the transition towards a discoid. Conversely, if material could be gently removed from the latter, it would keep its discoidal shape until the critical volume corresponds to $\alpha_1$, where it would dynamically evolve into a tactoid. The green lines in Fig.~\ref{fig:minimum_life_degeneratek31} delimit such a hysteresis loop. 

\begin{figure}[h]
	\centering
	\begin{subfigure}[c]{0.4\linewidth}
		\centering
		\includegraphics[width=0.7\linewidth]{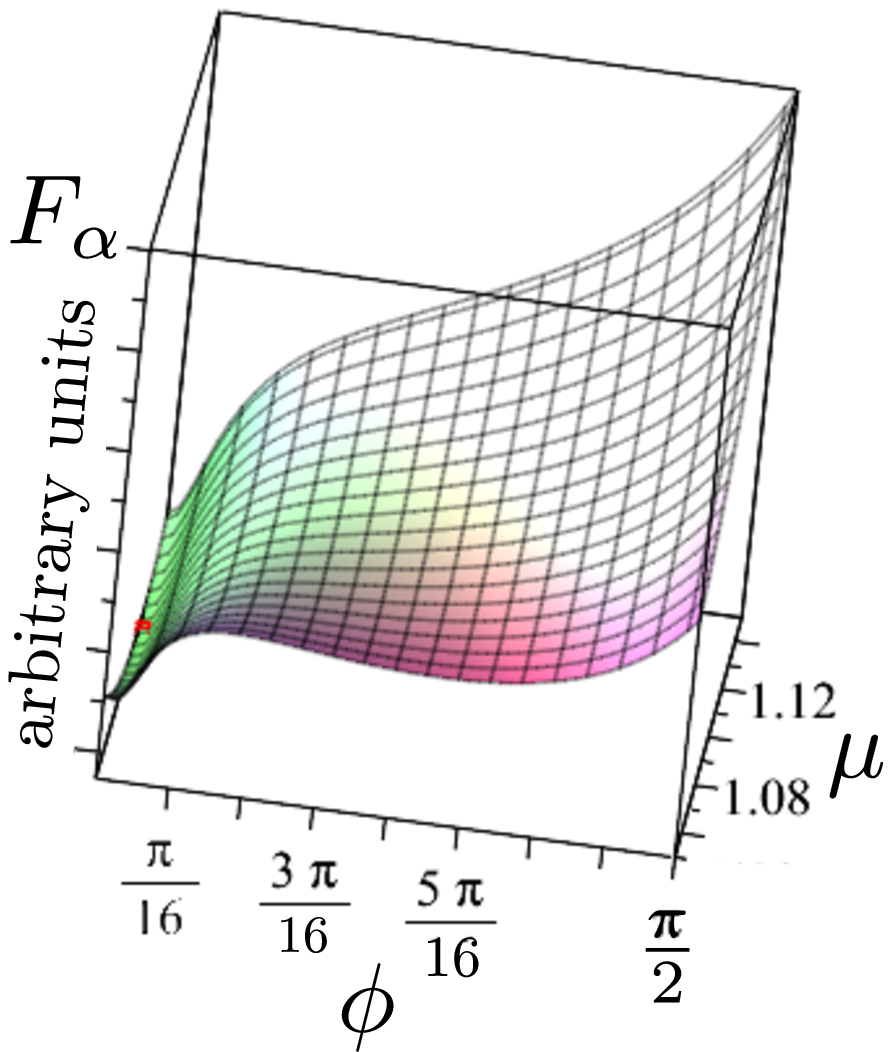}
		\caption{Graph of $\Fa$ against $\conf$; the red dot designates its single minimum.}
		\label{fig:minima1_a}
	\end{subfigure}
	\begin{subfigure}[c]{0.28\linewidth}
		\centering
		\includegraphics[width=.9\linewidth]{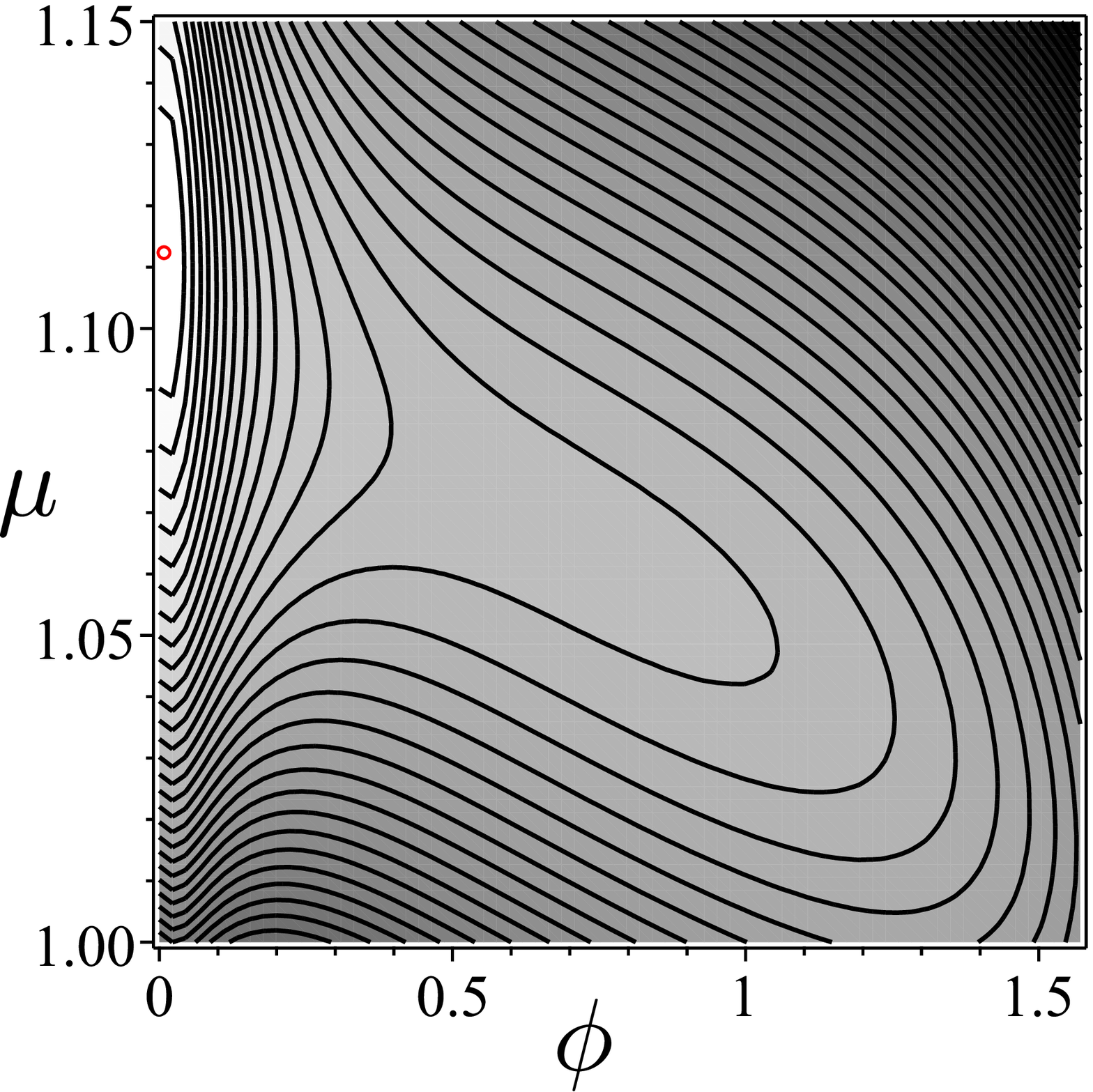}
		\caption{Contour plot of $\Fa$. The minimum is attained for $\phi\doteq0.01$ and $\mu\doteq1.11$ (red circle).}
		\label{fig:minima1_b}
	\end{subfigure}
	\begin{subfigure}[c]{0.3\linewidth}
		\centering
		\includegraphics[width=.5\linewidth]{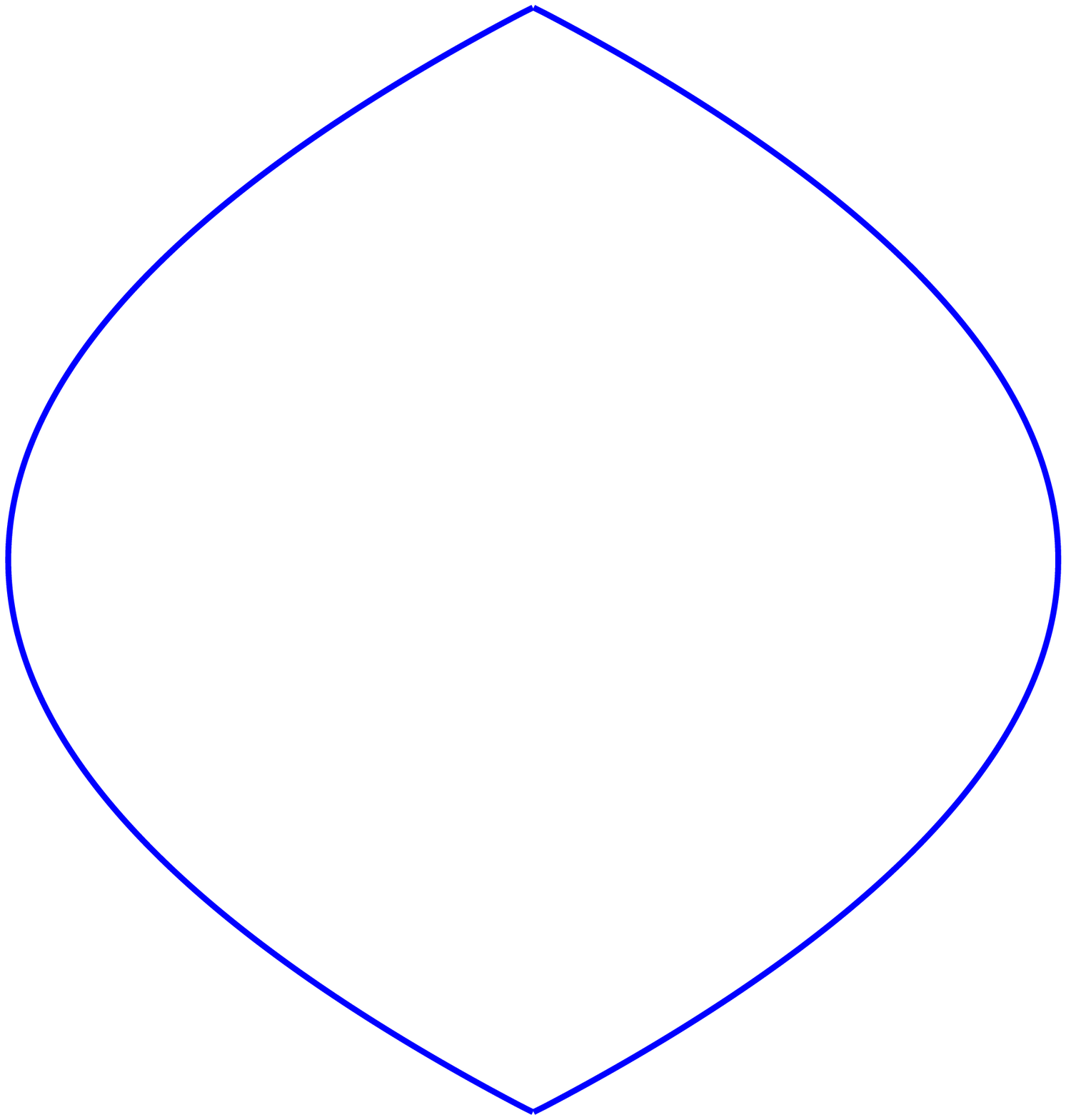}
		\caption{Equilibrium tactoidal shape corresponding through \eqref{eq:profile} to the minimizer of $\Fa$.}
		\label{fig:minima1_c}
	\end{subfigure}
	\caption{For $\alpha<\alpha_1$, the reduced free energy $\Fa$ is convex  on the configuration space $\conf$  and attains a single minimum in $0<\phi<\frac{\pi}{16}$. Here, $k_3=1$, $\alpha=170$, $\alpha_1=172$.}
	\label{fig:minima1}
\end{figure}

\begin{figure}[h]
	\centering
	\begin{subfigure}[c]{0.27\linewidth}
		\centering
		\includegraphics[width=1\linewidth]{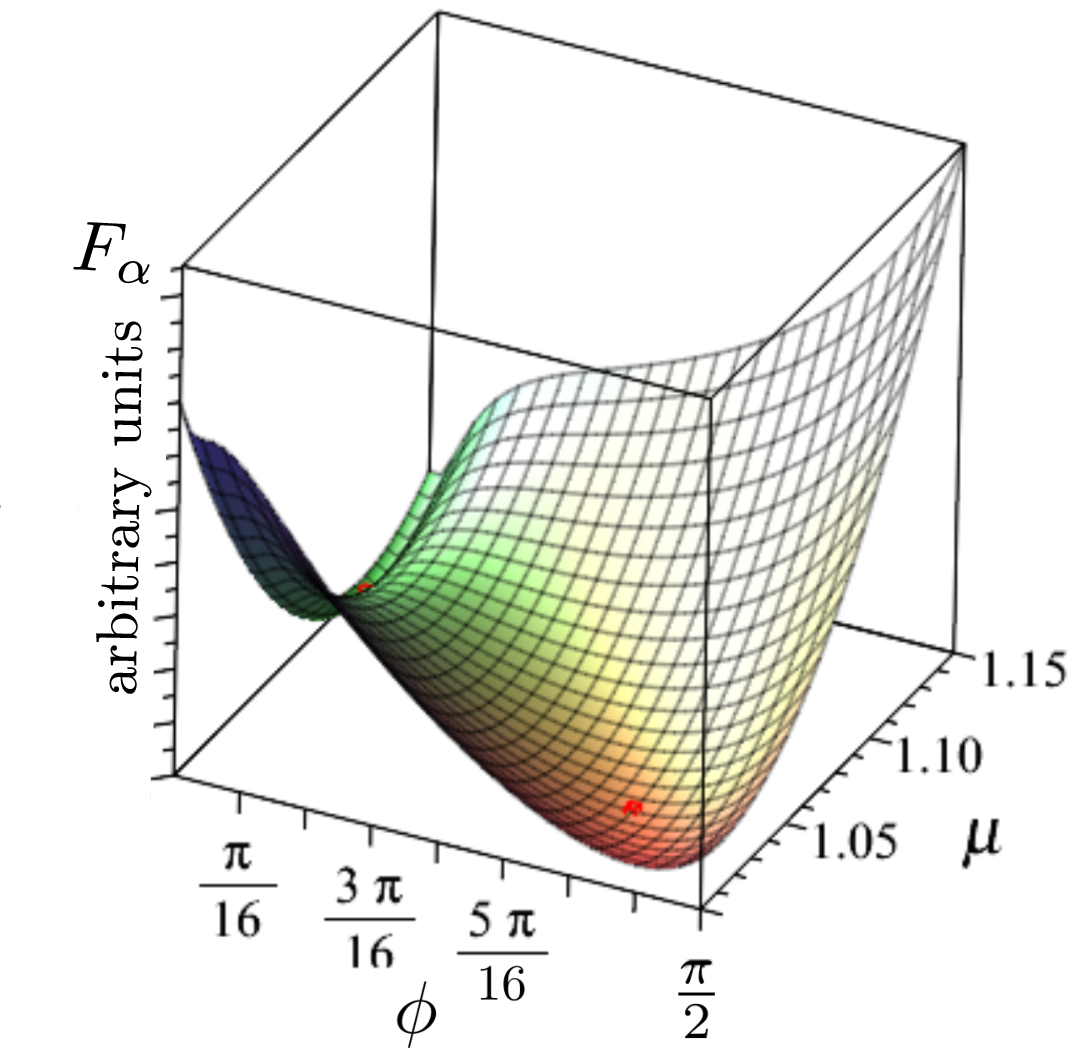}
		\caption{Graph of $\Fa$ against $\conf$. $\Fa$ is no longer convex: it attains two equal minima designated by red dots.}
		\label{fig:minima_bistability_a}
	\end{subfigure}
	\begin{subfigure}[c]{0.27\linewidth}
		\centering
		\includegraphics[width=.9\linewidth]{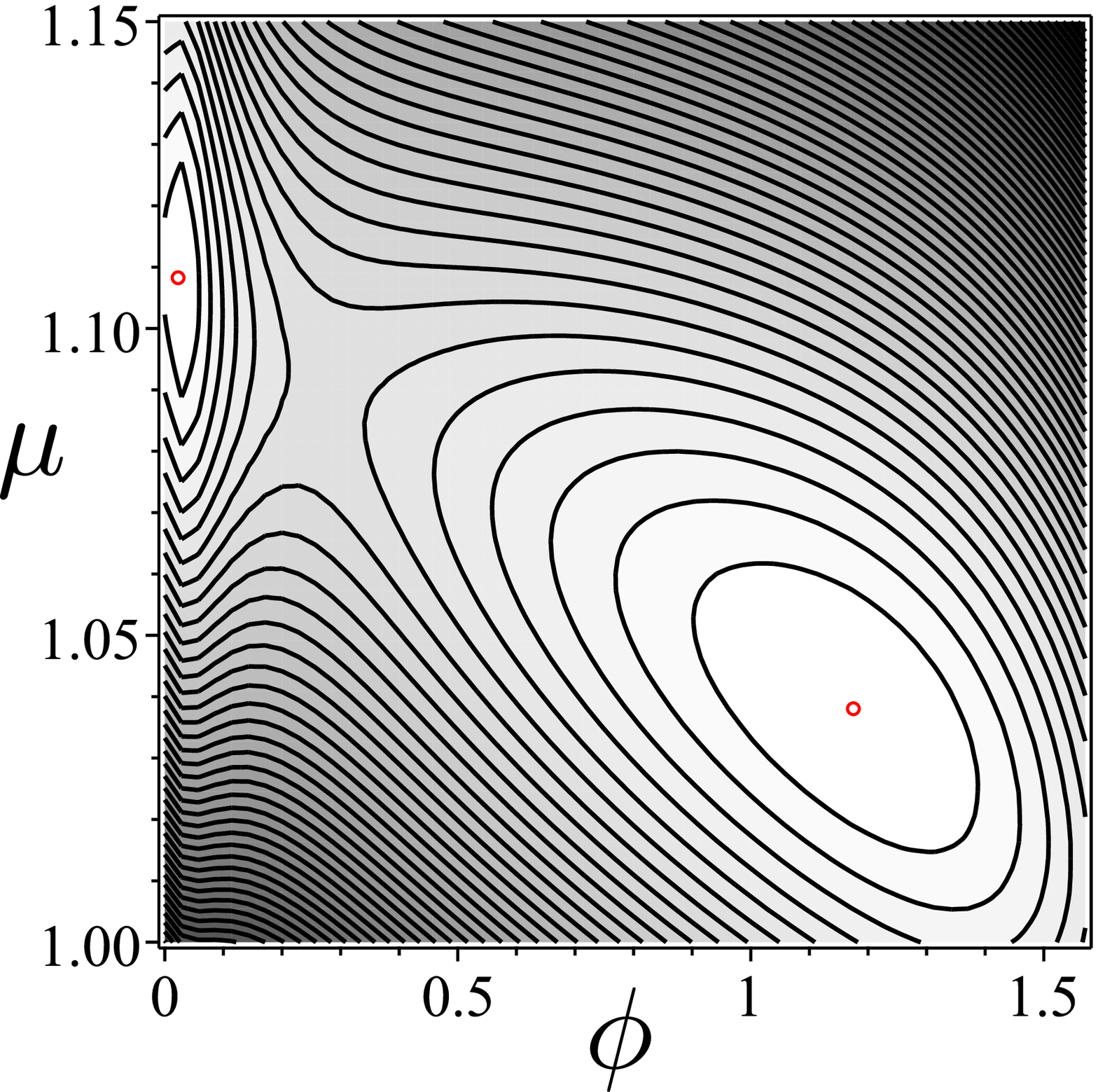}
		\caption{Contour plot of $\Fa$. One minimum is attained for $\phi\doteq0.025$ and $\mu\doteq1.11$ (tactoid), while the other is attained for $\phi\doteq1.2$ and $\mu\doteq1.04$ (doscoid); both are marked by red circles.}
		\label{fig:minima_bistability_b}
	\end{subfigure}
	\begin{subfigure}[c]{0.2\linewidth}
		\centering
		\includegraphics[width=0.7\linewidth]{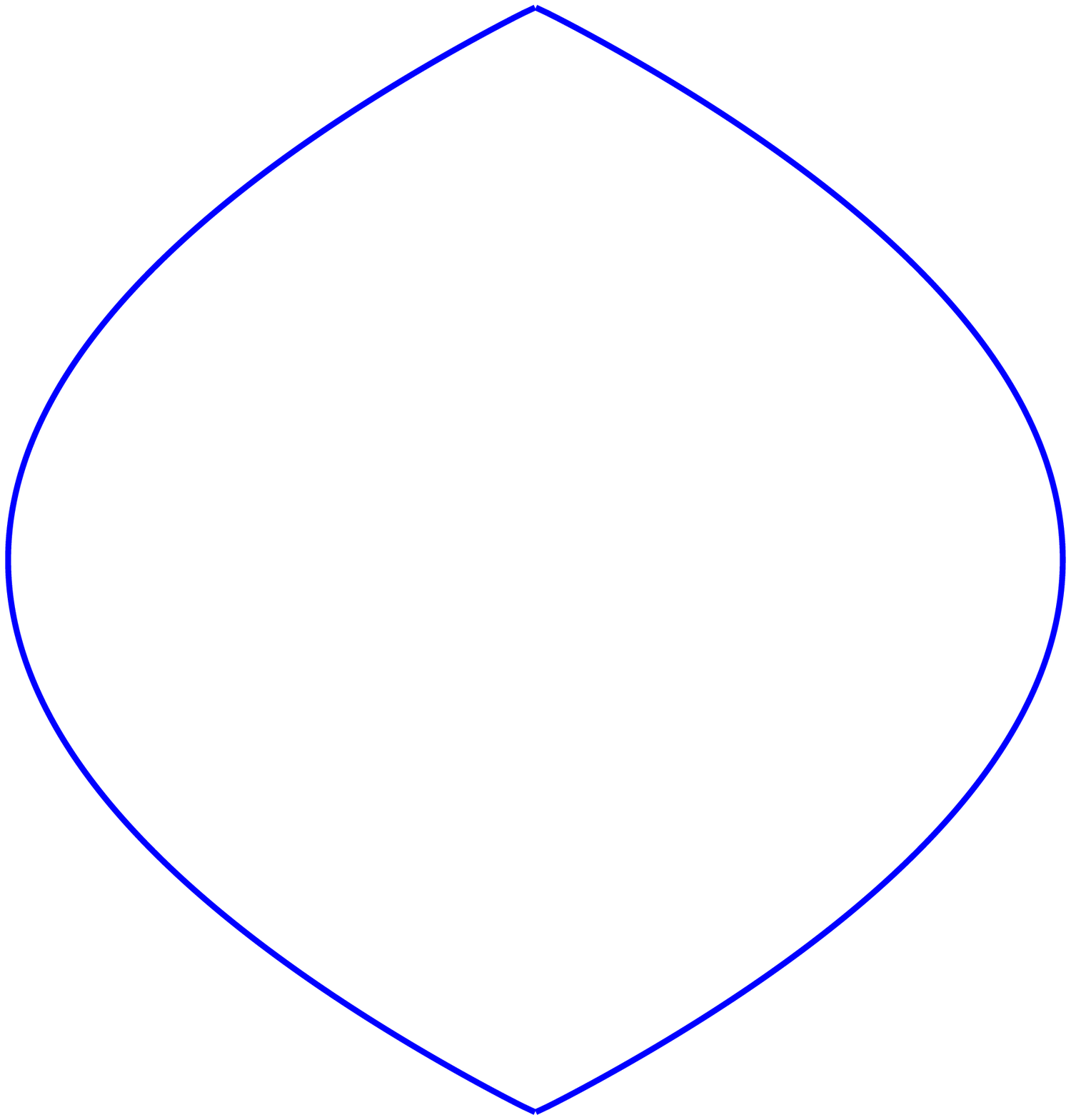}
		\caption{Tactoidal equilibrium shape.}
		\label{fig:minima_bistability_c}
	\end{subfigure}
	\begin{subfigure}[c]{0.2\linewidth}
		\centering
		\includegraphics[width=0.7\linewidth]{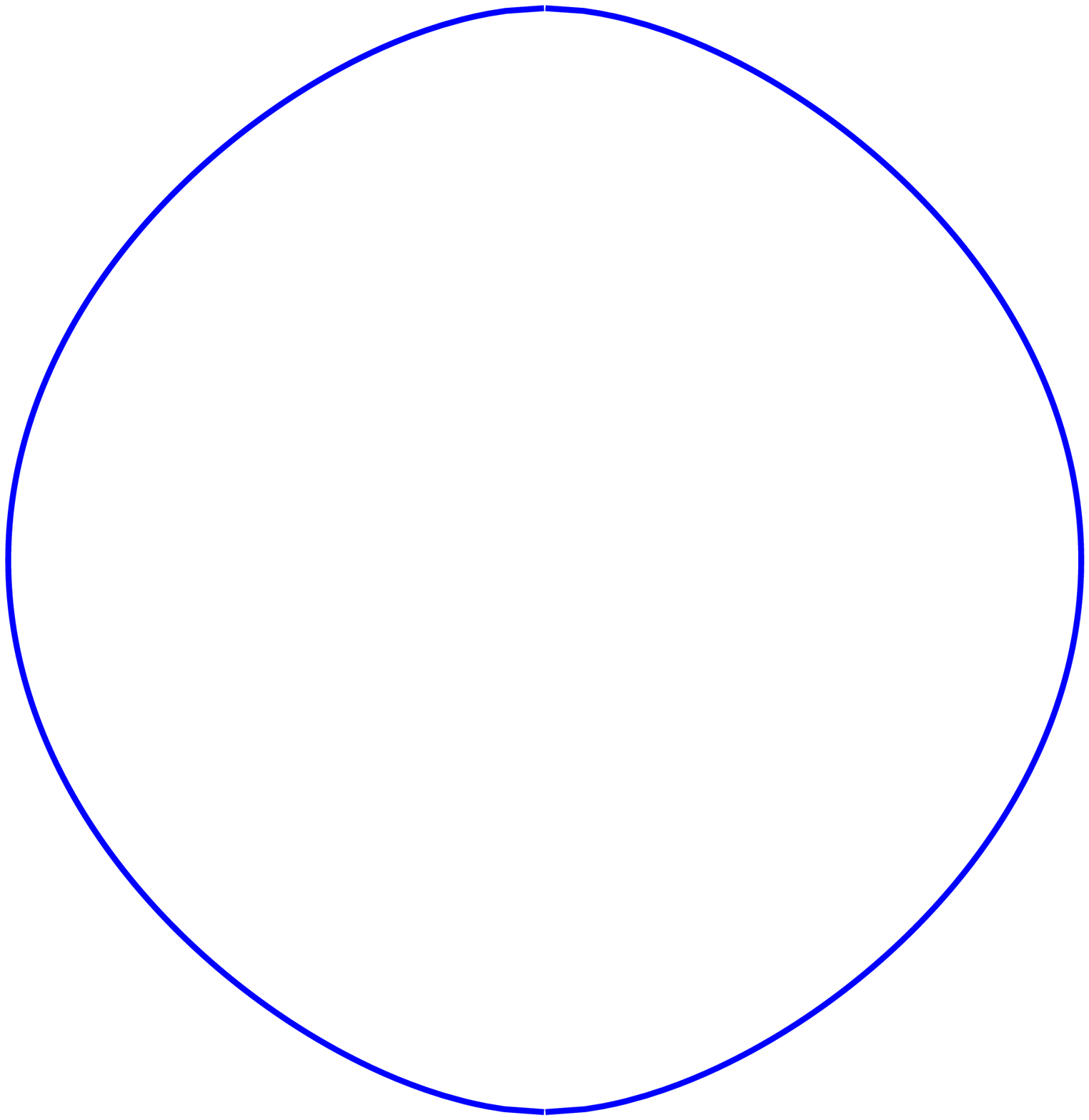}
		\caption{Discoidal equilibrium shape.}
		\label{fig:minima_bistability_d}
	\end{subfigure}
	\caption{For $\alpha=\alpha_\mathrm{b}$, the reduced free energy $\Fa$ has two (equal) global minima, one for $0<\phi<\frac{\pi}{16}$ (tactoid) and the other for  $\frac{\pi}{16}<\phi<\frac{\pi}{2}$ (discoid). Here, $k_3=1$ and $\alpha_\mathrm{b}=221$.}
	\label{fig:minima_bistability}
\end{figure}

Figure~\ref{fig:minima1} illustrates the energy landscape for $\alpha<\alpha_1$;
$\Fa$ is convex, and so it attains a single minimum, which falls in $0<\phi<\frac{\pi}{16}$, representing a tactoid. 
The scene changes in Fig.~\ref{fig:minima_bistability}, where $\alpha=\alpha_\mathrm{b}$ and $\Fa$ attains two equal minima, corresponding to a tactoid and a discoid.
For $\alpha>\alpha_2$, $\Fa$ is again convex, with a single minimum on a discoid,
\begin{figure}[h]
	\centering
	\begin{subfigure}[c]{0.4\linewidth}
		\centering
		\includegraphics[width=0.7\linewidth]{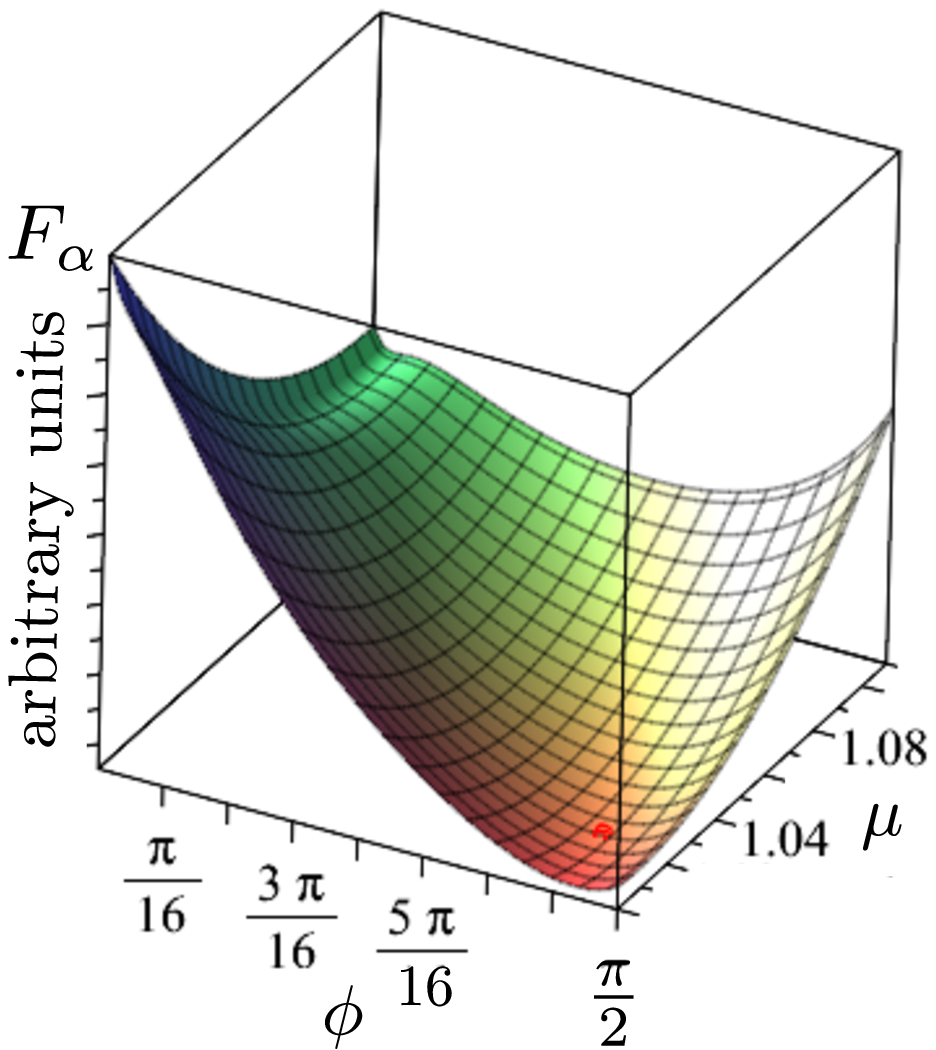}
		\caption{Graph of $\Fa$ against $\conf$; the red dot designates its single minimum.}
		\label{fig:minima3_a}
	\end{subfigure}
	\begin{subfigure}[c]{0.28\linewidth}
		\centering
		\includegraphics[width=.9\linewidth]{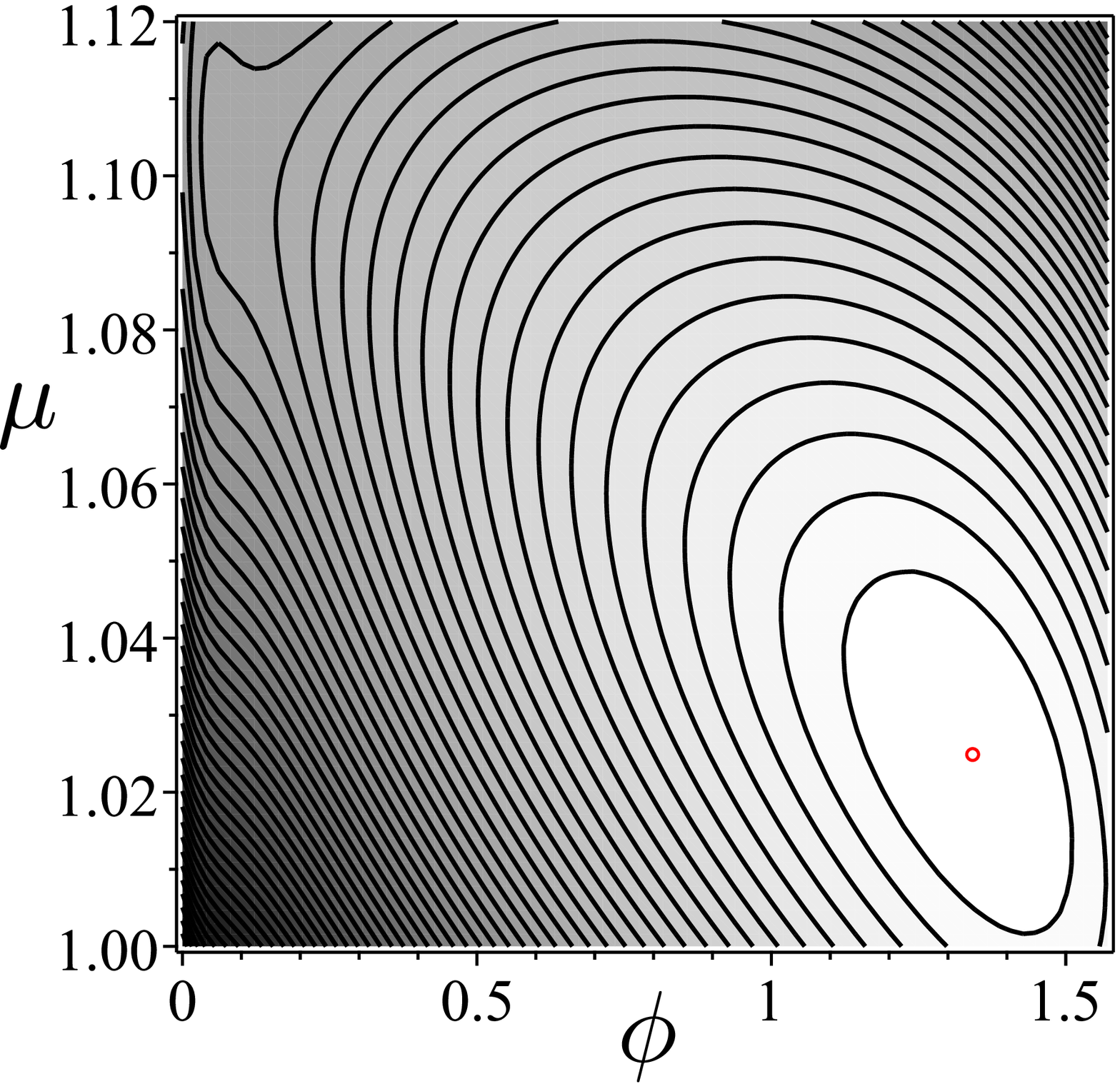}
		\caption{Contour plot of $\Fa$. The minimum is attained for $\phi\doteq1.34$ and $\mu\doteq1.02$ (red circle).}
		\label{fig:minima3_b}
	\end{subfigure}
	\begin{subfigure}[c]{0.3\linewidth}
		\centering
		\includegraphics[width=.5\linewidth]{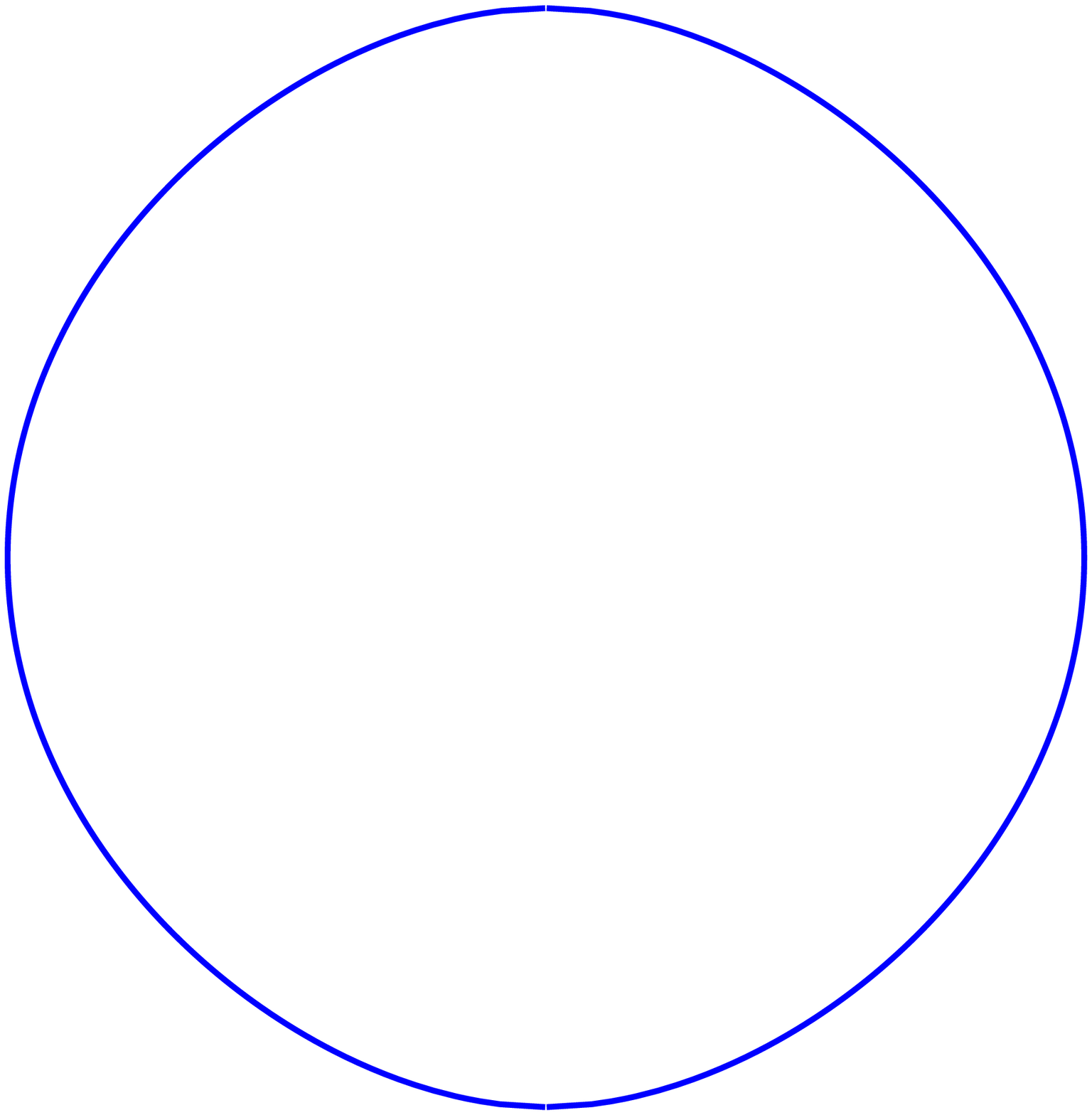}
		\caption{Equilibrium discoidal shape corresponding through \eqref{eq:profile} to the minimizer of $\Fa$.}
		\label{fig:minima3_c}
	\end{subfigure}
\caption{For $\alpha>\alpha_2$, the reduced free energy $\Fa$ is gain  convex  on the configuration space $\conf$ and attains a single minimum for $\frac{\pi}{16}<\phi<\frac{\pi}{2}$. Here, $k_3=1$, $\alpha=307$, $\alpha_2=305$.}
\label{fig:minima3}
\end{figure}
which gets closer and closer to the round disc, represented by the point $(\frac{\pi}{2},1)$ in $\conf$, as $\alpha$ grows indefinitely.

\begin{figure}[h] 
\includegraphics[width=.35\linewidth]{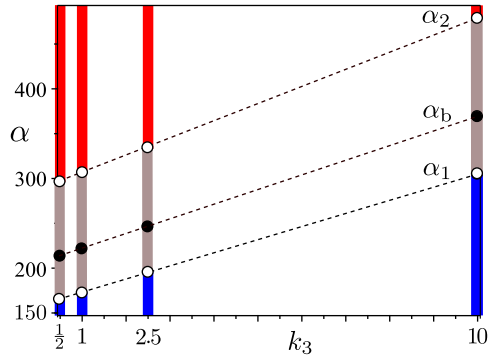}
\caption{Critical values of $\alpha$ for different values of $k_3$. To within a good approximation, they all depend linearly on $k_3$. Red and blue ranges refer to discoids and tactoids, respectively, according to the same color coding adopted in Fig.~\ref{fig:minimum_life_degeneratek31}.The broken lines represent the best linear fits in \eqref{eq:best_linear_fits}.}
\label{fig:hystogramk_3}
\end{figure}

We have computed the three critical values, $\alpha_1$, $\alpha_\mathrm{b}$, and $\alpha_2$ for several values of $k_3$. Fig.~\ref{fig:hystogramk_3} shows that, to within a good approximation, they all grow linearly  with $k_3$. 
The best linear fit is provided by the following functions,
\begin{subequations}\label{eq:best_linear_fits}
\begin{eqnarray}
\alpha_1&\approx157 +15 k_3,\\
\alpha_\mathrm{b}&\approx205 +16 k_3,\\	
\alpha_2&\approx289 +19 k_3.	
\end{eqnarray}
\end{subequations}

\section{Comparison with experiments}\label{sec:experiments_comparison}
A major motivation for this paper was offered by  the experiments conducted in \cite{kim:morphogenesis} with water solutions of disodium cromoglycate (DSCG). Here we apply our theory to interpret those experiments and show how to extract from them an estimate for the isotropic surface tension $\gamma$ at the interface between the nematic phase of a DSCG sol at a given concentration and its isotropic melt.

Among many other things, Kim et al.~\cite{kim:morphogenesis} explored  water solution of DSCG at concentration $c=16\,\mathrm{wt\%}$ confined between two parallel glass plates at a distance ranging in the interval
$1$-$5\,\mu \mathrm{m}$; the plates were  spin-coated with a polymide layer, SE-$7511$, which exerts a degenerate tangential anchoring on the nematic director.   This system can be treated as two-dimensional since the bounding plates suppress the out-of-plane distortions in the observed samples, as required by our theory.

Upon quenching the system from the isotropic phase into the coexistence regime, tactoidal droplets were observed, surrounded  by the parent isotropic phase. In particular (see, Fig.~6a of \cite{kim:morphogenesis}), a tactoid with bipolar nematic orientation was observed more closely, for which the area  $A_0=200\,\mu \mathrm{m}^2$ and the aspect ratio $\delta=1.3\pm 1$ were measured at the temperature $T=37.5\,^\circ\mathrm{C}$, which we read off from Fig.~2c of \cite{kim:morphogenesis}. We wish to make use of these data to validate our theory. To this end, we need to estimate the elastic constants $K_{11}$ and $K_{33}$ at $T=37.5\,^\circ\mathrm{C}$.
	
\begin{figure}[h]
	\centering
		\includegraphics[width=.4\linewidth]{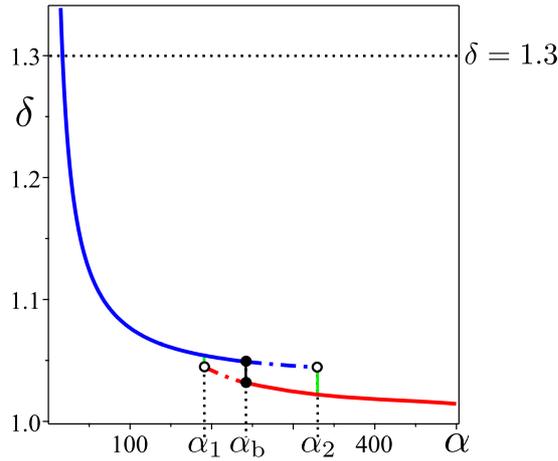}
		\caption{The droplet's aspect ratio $\delta$ computed on the minima of $\Fa$, plotted against $\alpha$ for $k_3=2.25$. The broken horizontal line designates the level $\delta=1.3$, which is attained for $\alpha=17.2$. Local and global minima are marked with the same symbols as in Fig.~\ref{fig:minimum_life_degeneratek31}. The critical values of $\alpha$ are $\alpha_1=192$, $\alpha_2=330$, and $\alpha_\mathrm{b}=243$. Equilibrium tactoids are genuine for $\alpha\lessapprox137$. }
		\label{fig:aspect_ratio_Kim}
	\end{figure}

Interpolating the curves representing in \cite{zhou:elasticity_2014}  the temperature dependence of the elastic constants of the nematic phase of DSCG at $c=16\,\mathrm{wt\%}$, we readily arrived at 
\begin{equation}
\label{eq:constants_Kim}
K_ {11} = 4\,\mathrm{pN}, \quad K_ {33} = 9\,\mathrm{pN}, \quad \hbox{at} \quad T=37.5\,^\circ\!\mathrm{C} \quad \hbox{and} \quad c=16\,\mathrm{wt\%},
\end{equation}
and so we take $k_3=2.25$. We can also obtain $\alpha$ from the measured aspect ratio $\delta$. It follows from \eqref{eq:profile} that in our theory $\delta$ can be given the form
\begin{equation}
\label{eq:aspect_ratio}
\delta:=\frac{\mu}{R(0)}=\frac{\mu^2\left(\frac83\cos\phi+\pi\sin\phi\right)}{\pi(\cos\phi+\sin\phi)},
\end{equation}
where use has also been made of \eqref{eq:a_b_representation}. The plot of $\delta$ computed on the minima of $\Fa$ over $\conf$ is drawn against $\alpha$ in Fig.~\ref{fig:aspect_ratio_Kim} for $k_3\doteq2.25$.
	We see that the value $\delta=1.3$ is attained by the tactoidal branch (as expected) for $\alpha\doteq17.2$; the corresponding coordinates in $\conf$ of the minimum of $\Fa$ are $\phi\doteq0$ and $\mu\doteq1.24$, and so the equilibrium shape is a genuine tactoid (with pointed tips). For $k_3=2.25$, equilibrium tactoids cease to be genuine at $\alpha\approx137$.

\begin{figure}[h]
	\centering
	\includegraphics[width=.6\linewidth]{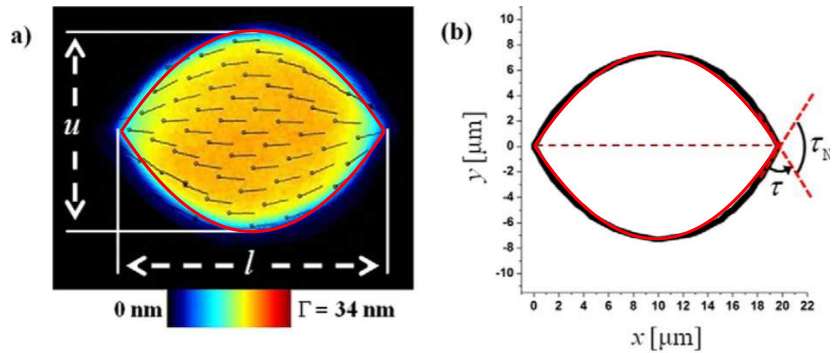}
	\caption{The droplet's equilibrium  shape (in red) corresponding to $\alpha=17.2$, $\phi=0$, and $\mu=1.24$ is contrasted against the shape observed in \cite{kim:morphogenesis} (see their Figs.~6a and 6b). Our estimate for the long axis is $l\approx20\,\mu\mathrm{m}$. On the left: PolScope image. On the right: reconstructed shape.}
	\label{fig:Comparison_Kim}
\end{figure}

Since the equivalent radius $\Req$ is known from the isoperimetric constraint \eqref{eq:area_constraintA0}, $\Req\approx8\mu\mathrm{m}$, we readily obtain for the long axis $l$ of the droplet $l=2R_0=2\mu\Req\approx20\,\mu\mathrm{m}$. The equilibrium tactoid predicted by our theory is plotted in Fig.~\ref{fig:Comparison_Kim} against the observed shape.
Our expected value for $l$ also agrees quite well with the measurement performed in \cite{kim:morphogenesis} on the reconstructed shape, as does the \emph{cusp angle} $\tau_\mathrm{N}$ marked in Fig.~\ref{fig:Comparison_Kim}. In our formalism, for a genuine tactoid, the latter can be expressed as 
\begin{equation}\label{eq:cusp_angle}
\tau_\mathrm{N}=-2\arctan(R'(\mu))=2\arctan\left(\frac{3\pi}{4\mu^2}\right),	
\end{equation}
which for $\mu=1.24$ delivers $\tau_\mathrm{N}\doteq1.99$. The value of $\tau_\mathrm{N}$ reported  in \cite{kim:morphogenesis} for the shape in Fig.~\ref{fig:Comparison_Kim} is $\tau_\mathrm{N}=1.05\pm0.05$, whereas a direct measurement on Fig.~\ref{fig:Comparison_Kim} would suggest that this value is indeed $\tau_\mathrm{N}/2$, in good agreement with our theoretical prediction. Moreover, by use of equation \eqref{eq:alpha}, we can give the following estimate for the  surface tension at the nematic/isotropic interface of an aqueous DSCG solution at $16\,\mathrm{wt\%}$,
\begin{equation}
\label{eq:gamma_Kim}
 \gamma\approx 0.89 \times 10^{- 5}\, \mathrm{J}/\mathrm{m}^2.
\end{equation}

Different, discordant estimates have been given in the literature for the order of magnitude of $\gamma$. For example, in \cite{mushenheim2014:using} they estimate $\gamma\sim10^{-6}\,\mathrm{J}/\mathrm{m}^2$, whereas  in \cite{kim:morphogenesis} they give $\gamma\sim10^{-4}\,\mathrm{J}/\mathrm{m}^2$, an estimate obtained by applying the pendant drop technique \cite{chen:interfacial,chen:interfacial_1999}. We trust that the measurements based on the theory presented in this paper might be more accurate. The value in \eqref{eq:gamma_Kim} is closer to that found in \cite{faetti:anchoring} at the nematic-isotropic interface of 5CB (see also \cite{faetti:measurements,faetti:nematic}).

The critical values of $\alpha$ corresponding to $k_3=2.25$ are $\alpha_1=192$, $\alpha_2=330$, and $\alpha_\mathrm{b}=243$, while the observed droplet shown in Fig.~\ref{fig:Comparison_Kim}  corresponds to  $\alpha=17.2$. According to our theory, one would then expect coexistence of tactoids and discoids for $\Req$ in the range $90\,\mu\mathrm{m}\lessapprox \Req\lessapprox 153\,\mu\mathrm{m}$, that is, for area $A_0$ in the range $0.3\,\mathrm{mm}^2\lessapprox A_0\lessapprox 0.7\,\mathrm{mm}^2$, a regime of large drops, for which no data are available in \cite{kim:morphogenesis}.

\section{Conclusion}\label{sec:conclusion}
We studied chromonic droplets in two space dimensions. Our motivation was the thorough experimental investigation in \cite{kim:morphogenesis} on thin cells with degenerate tangential anchoring for the nematic director. In particular, a solution of DSCG in water was quenched in the temperature regime where nematic and isotropic phases coexist in equilibrium, the former forming islands with a tactoidal shape.

We introduced a wide class of two-dimensional shapes for the equilibrium droplets, which includes tactoids, discoids, b\^atonnet, and butterflies, the latter of which are concave, whereas the former three are convex. Our analysis revealed that upon increasing the droplet's area a tactoidal equilibrium branch  gives way to a discoid branch, while neither   b\^atonnet nor butterflies can ever be equilibrium shapes. Moreover, there is a regime of shape coexistence, where a tactoid and a discoid are both local minima of the free energy, the global minimum shifting from one to the other at a critical value of the droplet's area, where perfect bistability is established. A typical bifurcation diagram with hysteresis describes the situation, reminiscent of a first-order phase transition with super-heating and super-cooling temperatures (replaced here by corresponding values of the area).

We put our theory to the test by interpreting experimental data provided by \cite{kim:morphogenesis}. We found a fairly good quantitative agreement between experiment and theory, although further data should be collected  to establish on firmer grounds the degree of confidence of the theory. In particular, we could extract from the data available in \cite{kim:morphogenesis} an estimate for the isotropic component $\gamma$ of the surface tension at the interface between coexisting nematic and isotropic phases of DSCG in  $16\,\%\mathrm{wt}$ aqueous sol. This estimate seems to promise more accuracy  than the rough evaluation of order of magnitude available in the literature for this material and its chromonic siblings. We hope that the theory proposed here could be used for a systematic determination of $\gamma$ for different temperatures and concentrations.

The available data do not cover the range of predicted bistability. We estimated the area that a droplet should reach to display an abrupt transition from tactoid to discoid.  It remains to be seen whether a controlled growth in the droplet's size can be realized to observe neatly this transition. 

We have shown that the critical values of the area that delimit the shape hysteresis are (increasing) linear functions of the ratio $K_{33}/K_{11}$ between bend and splay elastic constants. If this critical phenomenon  could be explored experimentally, our theory would also offer an independent way to measure $K_{33}/K_{11}$. This might be especially welcome for non conventional new lyotropic phases, such as chromonics.

\appendix
\section{Retracted Tangential Field}\label{sec:retracted_field}
In this Appendix, we justify the expression for the free-energy functional \eqref{eq:F_energy_functional} associated with the bipolar director field $\n$ defined as the unit vector field tangent to the lines with given $t$ and varying $y$ (see Fig.~\ref{fig:shape_b}). A generic curve in that family is represented by the position vector
\begin{equation}
\label{eq:curve_p_t}
\bm{p}_t(y):=g(t)R(y)\e_x+y\e_y, \quad-R_0\leqq y\leqq R_0,
\end{equation}
where $g$ is any function of class $C^1$ strictly increasing on $[0,1]$ and such that $g(0)=0$ and $g(1)=1$. 
For given $t\in[0,1]$, the tangent vector field $\n$ is given by 
\begin{eqnarray}
\label{eq:n}
\n=\frac{gR'\e_x+\e_y}{\sqrt{1+(gR')^2}},
\end{eqnarray}
where a prime denotes differentiation.
The unit vector field
\begin{equation}
\label{eq:n_perp}
\nper:=\e_z\times\n=\frac{-\e_x+gR'\e_y}{\sqrt{1+(gR')^2}}
\end{equation}
is everywhere orthogonal to $\n$ and such that $\nper|_{\boundaryR}=-\normal$, where $\normal$ is the outer unit normal to $\boundaryR$ (see again Fig.~\ref{fig:shape_b}).

Imagine now  a smooth curve in $\region$ parameterized as $\xi\mapsto(t(\xi),y(\xi))$; it follows from \eqref{eq:curve_p_t} and \eqref{eq:n} that
\begin{equation}
\label{eq:dot_curve_t}
\dot{\p}=g'R\dot{t}\e_x+\dot{y}(gR'\e_x+\e_y)=g'R\dot{t}\e_x+\sqrt{1+(gR')^2}\dot{y}\n,
\end{equation}
where a superimposed dot denotes specifically differentiation with respect to $\xi$. Thus,
the elementary area $\dd A$ is 
\begin{equation}
\label{eq:dA}
\dd A=\dd t\dd y g'R\sqrt{1+(gR')^2}\e_x\times\n\cdot\e_z =g'R\dd t\dd y,
\end{equation}
and the elementary length $\dd \ell_t$ on the retracted  curve $\boundaryR_t$, for given $t$, is
\begin{equation}
\label{eq:dl}
\dd \ell_t=\sqrt{1+(gR')^2}\dd y.
\end{equation}
In particular, the area of  $\region$ and the length of its boundary are given by
\begin{equation}
\label{eq:area}
A(\region)=2\int_{0}^{1}g'\dd t \int_{-R_0}^{R_0}R\dd y =2\int_{-R_0}^{R_0}R\dd y
\end{equation}
and 
\begin{equation}
\label{eq:length}
\ell(\boundaryR)=2\int_{-R_0}^{R_0}\sqrt{1+R'^2}\dd y.
\end{equation}

Differentiating $\n$ in \eqref{eq:n} along the smooth curve $\xi\mapsto(t(\xi),y(\xi))$, we find that  
\begin{equation}
\label{eq:dot_n}
\dot{\n}=-\frac{g'R'\dot{t}+gR''\dot{y}}{1+(gR')^2}\nper.
\end{equation}
Assuming that $\n$ is differentiable in $\region$, $\dot{\n}$ and $\dot{\p}$ must be related through
\begin{equation}
\label{eq:dot_n_=}
\dot{\n}=(\nabla\n)\dot{\p}.
\end{equation}
Since $\n$ is a unit vector field, $(\nabla\n)^\mathsf{T}$ annihilates $\n$, and so there exists a vector $\bm{a}=a_1\n+a_2\nper$ such that  
\begin{equation}
\label{eq:nabla_n_representation}
\nabla\n=\nper\otimes\bm{a}.
\end{equation}
To determine the scalar components $a_i$ of $\bm{a}$, we observe that, by \eqref{eq:nabla_n_representation}, \eqref{eq:dot_n_=} also reads as
\begin{equation}
\label{eq:dot_n_==}
\dot{\n}=(\bm{a}\cdot\dot{\p})\nper.
\end{equation}
Making use of \eqref{eq:dot_curve_t} and both \eqref{eq:n} and \eqref{eq:n_perp}, we readily see that 
\begin{equation}\label{eq:x_dotted_with_dot_p}
\bm{a}\cdot\dot{\p}=\frac{g'R(gR'a_1+a_2)\dot{t}}{\sqrt{1+(gR')^2}}+\sqrt{1+(gR')^2}a_1\dot{y},
\end{equation}
and inserting this  into \eqref{eq:dot_n_==} alongside with \eqref{eq:dot_n}, we obtain an identity for arbitrary $(\dot{t},\dot{y})$ only if
\begin{equation}\label{eq:a_b_components}
a_1=-\frac{gR''}{[1+(gR')^2]^{3/2}},\quad a_2=\frac{R'}{R}\frac{1}{\sqrt{1+(gR')^2}}-\frac{g^2R'R''}{[1+(gR')^2]^{3/2}},
\end{equation}
which leads us to 
\begin{equation}
\label{eq:nabla_n}
\nabla\n=-\frac{gR''}{[1+(gR')^2]^{3/2}}\nper\otimes\n
+\left(\frac{R'}{R}\frac{1}{\sqrt{1+(gR')^2}}-\frac{g^2R'R''}{[1+(gR')^2]^{3/2}}\right)\nper\otimes\nper.
\end{equation}
The following expressions for the traditional measures of distortion then follow from \eqref{eq:nabla_n},
\begin{subequations}\label{eq:distortion_measures}
\begin{align}
\diver\n&=\frac{R'}{\sqrt{1+(gR')^2}}\left(\frac{1}{R}-\frac{g^2R''}{1+(gR')^2}\right),\label{eq:div}\\
\curl\n&=-\frac{gR''}{[1+(gR')^2]^{3/2}}\e_z,\label{eq:curl}\\
\n&\cdot\curl\n=0,\label{eq:twist}\\
\n \times \curl\n&=\frac{gR''}{[1+(gR')^2]^{3/2}}\nper,\label{eq:bend}\\
\tr(\nabla\n)^2&-(\diver\n)^2=0.\label{eq:saddle_splay}
\end{align}	
\end{subequations} 

We found it useful to rescale all lengths to the radius $\Req$ of the disc of area $A_0$. Letting $\mu$ as in \eqref{eq:mu_definition} and using \eqref{eq:free_energy_functional} and \eqref{eq:area}, we arrive at the following reduced functional,
\begin{align}
\label{eq:F_energy_functional_t} 
&F[\mu;R]:=\frac{\free[\body]}{K_{11}h}=\nonumber\\
&=\int_{-\mu}^{\mu} \dd y\int_0^1 g'\left[\frac{g^4RR'^2R''^2}{\big(1+(gR')^2\big)^3}+\frac{R'^2}{R}\dfrac{1}{\big(1+(gR')^2\big)}-2\frac{g^2R'^2R''}{\big(1+(gR')^2\big)^2}+k_3\frac{g^2RR''^2}{\big(1+(gR')^2\big)^3}\right]\dd t+2\alpha\int_{-\mu}^{\mu}\sqrt{1+R'^2}\dd y,
\end{align}
where $k_3$ and $\alpha$ are as in \eqref{eq:elastic_constants_rescaled} and in \eqref{eq:alpha}. The integration in $t$, which delivers   \eqref{eq:F_energy_functional} in the main text, is independent of the specific function $g$, provided it is monotonic and obeys the prescribed boundary conditions.

\section{Preventing Anchoring Breaking}\label{sec:alpha_safeguard}
Here, we perform an energy comparison to identify the \emph{safeguard} value $\alpha_\mathrm{s}$ of $\alpha$, that is, the lower bound that should be exceeded for a drop to be bipolar at equilibrium. It is known that for $\alpha$ sufficiently small the tangential anchoring favored by the interfacial energy \eqref{eq:free_sup} is bound to be broken, so that the nematic alignment becomes uniform throughout the droplet. Uniform and bipolar alignments will indeed be the terms of comparison for our estimate of $\alpha_\mathrm{s}$. 

\begin{figure}[h]
	\centering
	\begin{subfigure}[c]{0.45\linewidth}
		\centering
		\includegraphics[width=.33\linewidth]{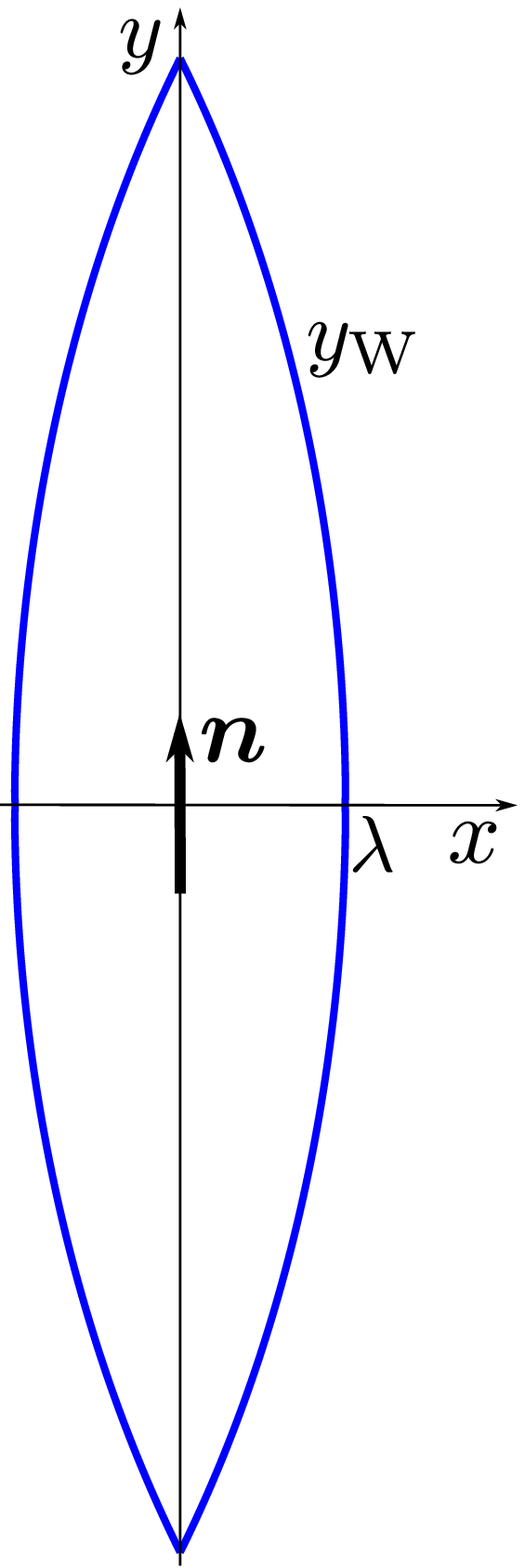}
		\caption{Wulffian shape obtained by solving equations \eqref{eq:Wulff_construction} for $\omega=5$. With lengths scaled to $\Req$,  $\lambda=0.51$.}
		\label{fig:Wulffian}
	\end{subfigure}
\qquad
	\begin{subfigure}[c]{0.45\linewidth}
		\centering
		\includegraphics[width=\linewidth]{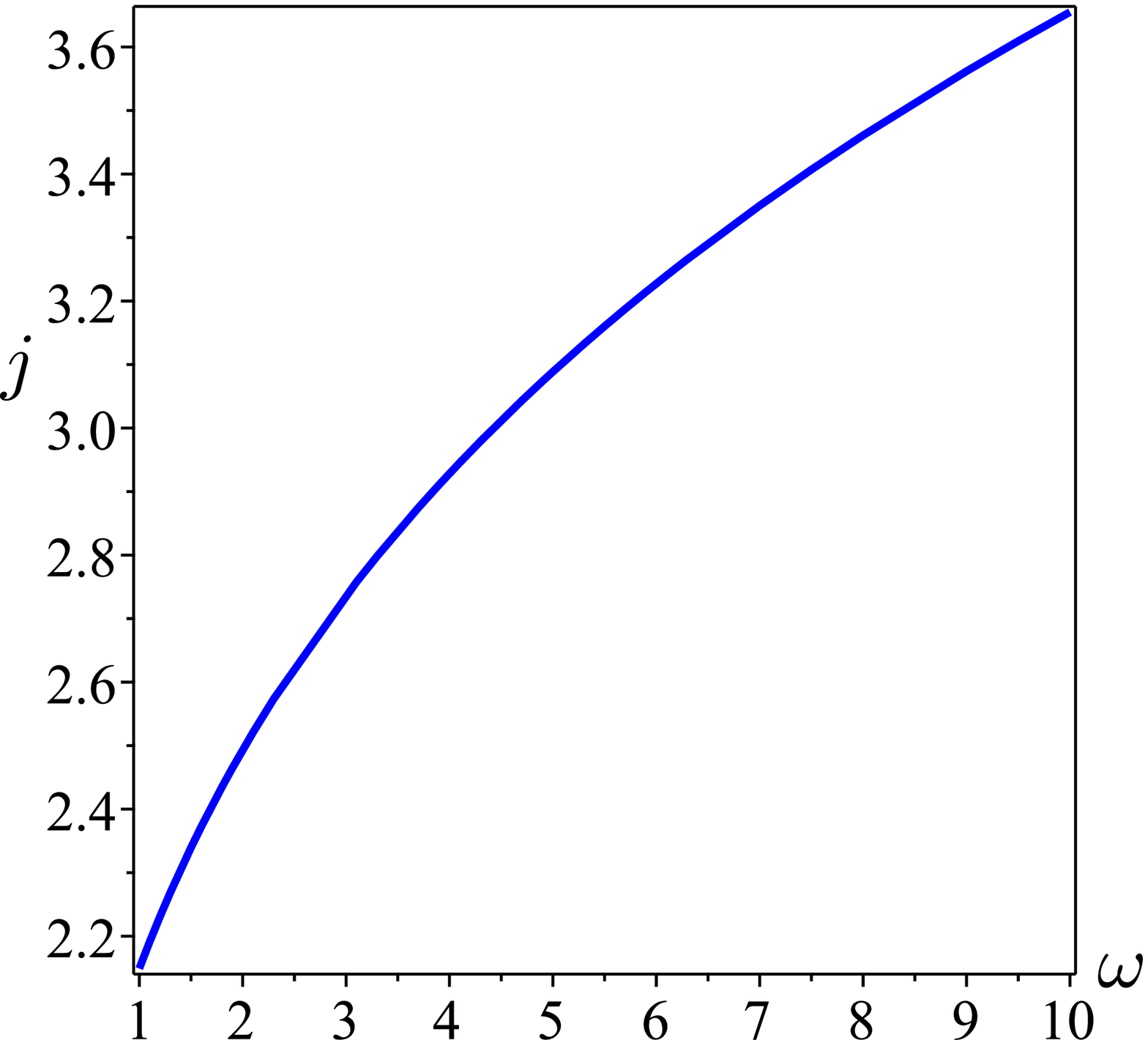}
		\caption{Graph of the function $j$ in \eqref{eq:j_definition} plotted against $\omega$.}
		\label{fig:Wulffian}
	\end{subfigure}
	\caption{Shape and (scaled) energy for the fully aligned droplet.}
	\label{fig:Wulff_shape}
\end{figure}

For drops with uniform alignment, the total free energy reduces to  \eqref{eq:free_sup}, subject to the area constraint \eqref{eq:area_rescaled}. The optimal shape is delivered by the classical Wulff's construnction \cite{wulff:frage} (see also \cite[p.\,490]{kleman:soft}). Assuming that $\n\equiv\e_y$ and that $\region$ is mirror-symmetric with respect to both axes $\e_x$ and $\e_y$, one needs only determine the equilibrium shape of $\region$ in the positive $(x,y)$ quadrant. We let $y=y_\mathrm{W}(x)\geqq0$ represent the profile of $\region$ and define $\lambda>0$ by setting $y_\mathrm{W}(\lambda)=0$. Here, as in the main text, all lengths are scaled to $\Req$. The method illustrated in \cite[Ch.5]{virga:variational} and \cite{prinsen:shape} delivers Wulff's shape through the following explicit function  
\begin{subequations}\label{eq:Wulff_construction}
\begin{equation}
\label{eq:Wulff_profile}
y_\mathrm{W}\left(x\lambda^{-1}\right)\lambda^{-1}=\frac{1+\omega\left(1-\xi^2\right)-\xi x\lambda^{-1}}{\sqrt{1-\xi^2}}.
\end{equation}
where $\xi$ is given in terms of $x$ by solving the algebraic equation
\begin{equation}
\label{eq:xi_equation}
\omega \xi^3+(1-\omega)\xi-x\lambda^{-1}=0
\end{equation}
and $\lambda$ is determined by the isoperimetric constraint \eqref{eq:area_rescaled}, which here reads as 
\begin{equation}
\label{eq:area_basis_Wulff}
4\int_0^\lambda y_\mathrm{W}(x)\dd x = 4\lambda^2\int_0^1y_\mathrm{W}\left(x\lambda^{-1}\right)\lambda^{-1} \dd \left(x\lambda^{-1}\right)=\pi.
\end{equation} 
\end{subequations}
Correspondingly, by computing the free energy in \eqref{eq:free_sup} on Wulff's shape, we obtain that  
\begin{equation}
\label{eq:free_a_hom_degenerate}
\free_\mathrm{W}=4K_{11}h\alpha j(\omega),  
\end{equation}
where the function $j$, defined by
\begin{equation}
	\label{eq:j_definition}
	j(\omega):=\frac{1}{2}\sqrt{\frac{\pi}{\int_0^1y_\mathrm{W}\dd x}}\left(\int_0^1\sqrt{1+y_\mathrm{W}'^2}\dd x+\omega\int_0^1\frac{1}{\sqrt{1+y_\mathrm{W}'^2}}\dd x\right),
\end{equation}
is computed on the solution $y_\mathrm{W}$ of \eqref{eq:Wulff_construction}, renormalized so that $y_\mathrm{W}(1)=0$. Figure~\ref{fig:Wulff_shape} illustrates the Wulffian shape obtained with this method for $\omega=5$ and the graph of $j$ for $0\leqq\omega\leqq10$.

\begin{figure}[h]
	\centering
	\includegraphics[width=0.33\linewidth]{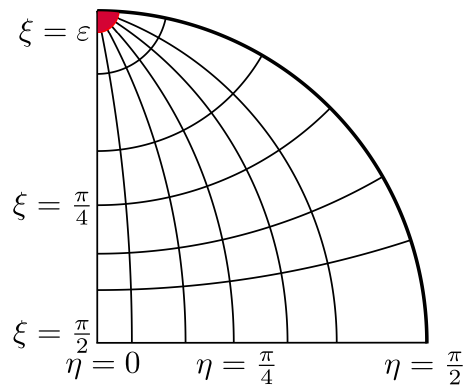}
	\caption{A quarter of a unit disc described in bipolar coordinates $(\xi,\eta)$. Apollonian circles correspond to different values of $\eta$; they are orthogonal to the coordinate lines with $\xi$ constant. As in Fig.~\ref{fig:cut}, also here $\vae$ is out of scale.}
	\label{fig:areoles_W_b}
\end{figure}

We now compare this energy to that of a disc with an in-plane bipolar director field whose integral lines are Apollonian circles passing through the poles  (see, for example, $\S$\,2 of \cite{ogilvy:excursions}); their  radius increases to $+\infty$ upon approaching the $y$-axis, as shown in Fig.~\ref{fig:areoles_W_b}, which represents a quadrant of the disc.
Adapting the computations in bipolar coordinates $(\xi,\eta)$ of Williams~\cite{williams:transitions} to the present two-dimensional setting, we arrive at
\begin{align}
\label{eq:F_energy_functional_Williams_reduced} 
\free_{\mathrm{bip}}&:=K_{11}h\left\{4\int_{\vae}^{\frac{\pi}{2}} \dd \xi\int_0^{\frac{\pi}{2}} \dd \eta \left[\frac{1}{2}\frac{\cos^2\xi}{\sin\xi}\frac{1}{(1+\sin\xi\cos\eta)^2}+\frac{k_3}{2}\frac{\sin^2\eta\sin\xi}{(1+\sin\xi\cos\eta)^2}\right]+2\alpha\pi\right\}\nonumber\\&=4hK_{11}\left\{\left[\frac{\pi}{4}\left(k_3-1-k_3\ln2-\ln\varepsilon\right)+\varepsilon+\alpha\left(\frac{\pi}{2}-\varepsilon\right)\right]+\frac{\alpha\omega \pi \varepsilon}{2}\right\},
\end{align}
where the detect core has been identified with the region $0\leqq\xi\leqq\vae$. The estimate in \eqref{eq:alpha_safeguard} follows from requiring that $\free_\mathrm{bip}<\free_\mathrm{W}$.

%

\end{document}